\documentclass[sigconf]{acmart}

\usepackage{enumitem}
\usepackage{colortbl}
\definecolor{mygray}{gray}{0.8}
\usepackage{flushend}
\usepackage{balance}
\usepackage{booktabs}
\usepackage{graphicx}
\usepackage{amsmath}
\usepackage{algorithm}
\usepackage{algorithmic}

\usepackage{xcolor}
\definecolor{deeppurple}{rgb}{0.40, 0.20, 0.60}

\usepackage{lipsum}
\usepackage{subfig}
\usepackage{color}

\usepackage{bigstrut}
\usepackage{multirow}
\usepackage{siunitx}
\usepackage{listings}
\usepackage{eso-pic}
\usepackage{tikz}
\usepackage{xspace}
\usepackage{fancyhdr}
\usepackage{mathtools}
\usepackage{soul}
\usepackage[leftcaption]{sidecap}
\usepackage{diagbox}
\usepackage{makecell}
\usepackage{hyperref}
\usepackage{listings}
\usepackage{xcolor} % 为代码上色
\AtBeginDocument{%
  }

\copyrightyear{2025} 
\acmYear{2025} 
\setcopyright{cc}
\setcctype{by}
\acmConference[ISCA '25]{Proceedings of the 52nd Annual International
Symposium on Computer Architecture}{June 21--25, 2025}{Tokyo, Japan}
\acmBooktitle{Proceedings of the 52nd Annual International Symposium on
Computer Architecture (ISCA '25), June 21--25, 2025, Tokyo,
Japan}\acmDOI{10.1145/3695053.3731099}
\acmISBN{979-8-4007-1261-6/2025/06}

\begin{document}

\title{DaDu-\textsc{Corki}\xspace: Algorithm-Architecture Co-Design for Embodied AI-powered Robotic Manipulation}

\author{Yiyang Huang}
\authornote{equal contribution.}
\authornote{part of this work was done during the first author's internship at Meituan.}
\affiliation{
  \institution{Institute of Computing Technology, Chinese Academy of Sciences \\ University of Chinese Academy of Sciences}
  % \streetaddress{P.O. Box 1212}
  \city{Beijing}
   \country{China}
}

\author{Yuhui Hao}
\authornotemark[1]
\affiliation{
  \institution{Institute of Computing Technology, Chinese Academy of Sciences \\ University of Chinese Academy of Sciences}
  % \streetaddress{P.O. Box 1212}
  \city{Beijing}
  \country{China}
}

\author{Bo Yu}
\affiliation{%
  \institution{Shenzhen Institute of Artificial Intelligence and Robotics for Society}
  \city{ShenZhen}
  %\city{Shenzhen}
   \country{China}
}

\author{Feng Yan}
\affiliation{%
  \institution{Meituan}
  \city{Beijing}
  \country{China}
}

\author{Yuxin Yang}
\affiliation{%
  \institution{Institute of Computing Technology, Chinese Academy of Sciences}
  % \streetaddress{30 Shuangqing Rd}
  \city{Beijing}
  % \state{Beijing Shi}
   \country{China}
  }

\author{Feng Min}
\affiliation{%
  \institution{Institute of Computing Technology, Chinese Academy of Sciences }
  % \streetaddress{30 Shuangqing Rd}
  \city{Beijing}
  % \state{Beijing Shi}
   \country{China}
  }
  
\author{Yinhe Han}
\affiliation{%
  \institution{Institute of Computing Technology, Chinese Academy of Sciences}
  % \streetaddress{30 Shuangqing Rd}
  \city{Beijing}
  % \state{Beijing Shi}
   \country{China}
  }

\author{Lin Ma}
\authornote{corresponding author, questions could be addressed to ganyiming@ict.ac.cn.}
\affiliation{%
  \institution{Meituan}
  % \streetaddress{30 Shuangqing Rd}
  \city{Beijing}
  % \state{Beijing Shi}
   \country{China}
  }

\author{Shaoshan Liu}
\authornotemark[3]
\affiliation{%
  \institution{Shenzhen Institute of Artificial Intelligence and Robotics for Society}
  % \streetaddress{1 Th{\o}rv{\"a}ld Circle}
  \city{Shenzhen}
   \country{China}
  % \thanks{$^*$ indicates equal contribution to the paper.}
  % \thanks{$\dagger$ indicates the corresponding author of the paper.}
  }

\author{Qiang Liu}
\affiliation{%
 \institution{Tianjin University}
 \city{Tianjin}  
  \country{China}
 }

\author{Yiming Gan}
\authornotemark[3]
\affiliation{%
  \institution{Institute of Computing Technology, Chinese Academy of Sciences}
  \city{Beijing}
   \country{China}
  }
\renewcommand{\shortauthors}{Huang and Hao et al.}

\begin{abstract}
Embodied AI robots have the potential to fundamentally improve the way human beings live and manufacture. Continued progress in the burgeoning field of using large language models to control robots depends critically on an efficient computing substrate, and this trend is strongly evident in manipulation tasks. In particular, today's computing systems for embodied AI robots for manipulation tasks are designed purely based on the interest of algorithm developers, where robot actions are divided into a discrete frame basis. Such an execution pipeline creates high latency and energy consumption. This paper proposes \textsc{Corki}\xspace, an algorithm-architecture co-design framework for real-time embodied AI-powered robotic manipulation applications. We aim to decouple LLM inference, robotic control, and data communication in the embodied AI robots' compute pipeline. Instead of predicting action for one single frame, \textsc{Corki}\xspace predicts the trajectory for the near future to reduce the frequency of LLM inference. The algorithm is coupled with a hardware that accelerates transforming trajectory into actual torque signals used to control robots and an execution pipeline that parallels data communication with computation. \textsc{Corki}\xspace largely reduces LLM inference frequency by up to $5.1\times$, resulting in up to $5.9\times$ speed up. The success rate improvement can be up to 13.9\%.

% \proj largely reduces LLM inference frequency by up to $8.0\times$, resulting in up to $9.1\times$ speed up. The success rate improvement can be up to 17.3\%. 

% Code is provided for re-implementation.
% \href{https://github.com/hyy0613/Corki}{https://github.com/hyy0613/Corki}

\end{abstract}



\received{20 February 2007}
\received[revised]{12 March 2009}
\received[accepted]{5 June 2009}

\maketitle

\section{Introduction}
\label{sec:intro}

Large Language Models (LLMs) have demonstrated remarkable capabilities in reasoning and long-term task planning \cite{RT1,RT2,driess2023palm,huang2023voxposer,duan2022survey,song2023llm,liu2024robouniviewvisuallanguagemodelunified}. Building upon the success of LLMs, the field of embodied AI, which employs LLMs to control robots interacting with the physical world, is increasingly recognized as a promising step towards achieving Artificial General Intelligence (AGI).

\begin{sloppypar}
The single most important difference between using LLMs for generating text and images versus integrating them as decision-making and planning modules within robotic pipelines lies in the hard real-time constraints imposed on robots \cite{mei2006deployment,khatib1986real}. Without real-time assurances, the applicability of embodied AI systems is severely limited to theoretical studies rather than real-world applications. 
\end{sloppypar}

In the robotics domain, there are two primary approaches to using LLMs. The first involves incorporating an LLM as the central decision-making module, where it decomposes complex tasks into fundamental robotic actions~\cite{saycan2022arxiv, liang2023code,vemprala2024chatgpt}. This high-level approach relies on the reasoning capabilities of LLMs and focuses on seamlessly integrating them into the robotics software pipeline. Here, the LLMs are typically very large in terms of parameters, can be single-modality, and are hosted on cloud servers (e.g., GPT-4). Given the low frequency of decision-making in robotic applications, the latency of communication and LLM inference can be tolerable. The second approach uses a smaller, usually multi-modality, LLM to guide basic robot control. This low-level case alters traditional program-based control and is highly sensitive to latency because the robot control module exercises a much higher frequency. Our work focuses on this second approach, and for the remainder of this paper, 'embodied AI systems' will refer to this direction.

While using LLMs to control high degree-of-freedoms (DoF) robots such as humanoid robots may still be a distant goal, embodied AI algorithms have transformed manipulation tasks. Currently, the mainstream work flow for a robot to perform manipulation tasks without using embodied AI algorithms relies on efforts from expert programmers, where
%Traditionally, for a robot to pick up an object from a table, 
the programmer must specify the starting position, target position, and approach angles. Any change in the object’s location requires reprogramming. With advances in embodied AI, however, a robot arm can now manipulate different objects from any location on the table using only cameras and human instructions~\cite{zhaolearning,livision, team2024octo, wu2023unleashing, o2023open, cheang2024gr2generativevideolanguageactionmodel}. This paper will focus on manipulation tasks.

Current embodied AI systems struggle to meet real-time constraints. The fundamental reason lies in the execution model of embodied AI systems. To date, almost all embodied AI systems follow a sequential execution model that processes video input and generates robot actions on a frame-by-frame basis~\cite{RT1, RT2, team2024octo,kim2024openvla,o2023open}. Specifically, after warming up, the robots will start with a video sequence containing $N$ frames and a language instruction $i$. The LLM will predict the robot action tuple $(\Delta{x}, \Delta{y}, \Delta{z}, ...)$ based on the current input tuple $(Frame_{t-N}, Frame_{t-N+1}, ... , Frame_{t}, i)$, where $\Delta$ denotes the proposed robot movements and $Frame_{t}$ represents images within the video sequence. Upon executing the action, the robot captures a subsequent frame $Frame_{t+1}$ at the latest position. The next LLM inference then involves processing the updated input tuple $(Frame_{t-N+1}, Frame_{t-N+2}, ... , Frame_{t+1}, i)$.

\begin{figure}[t]
\centering
\subfloat[The current discrete execution pipeline of embodied AI systems, where every time a single next step action is predicted and all three stages happen for every frame.]
{
  \includegraphics[trim=0 0 0 0, clip, width=\columnwidth]{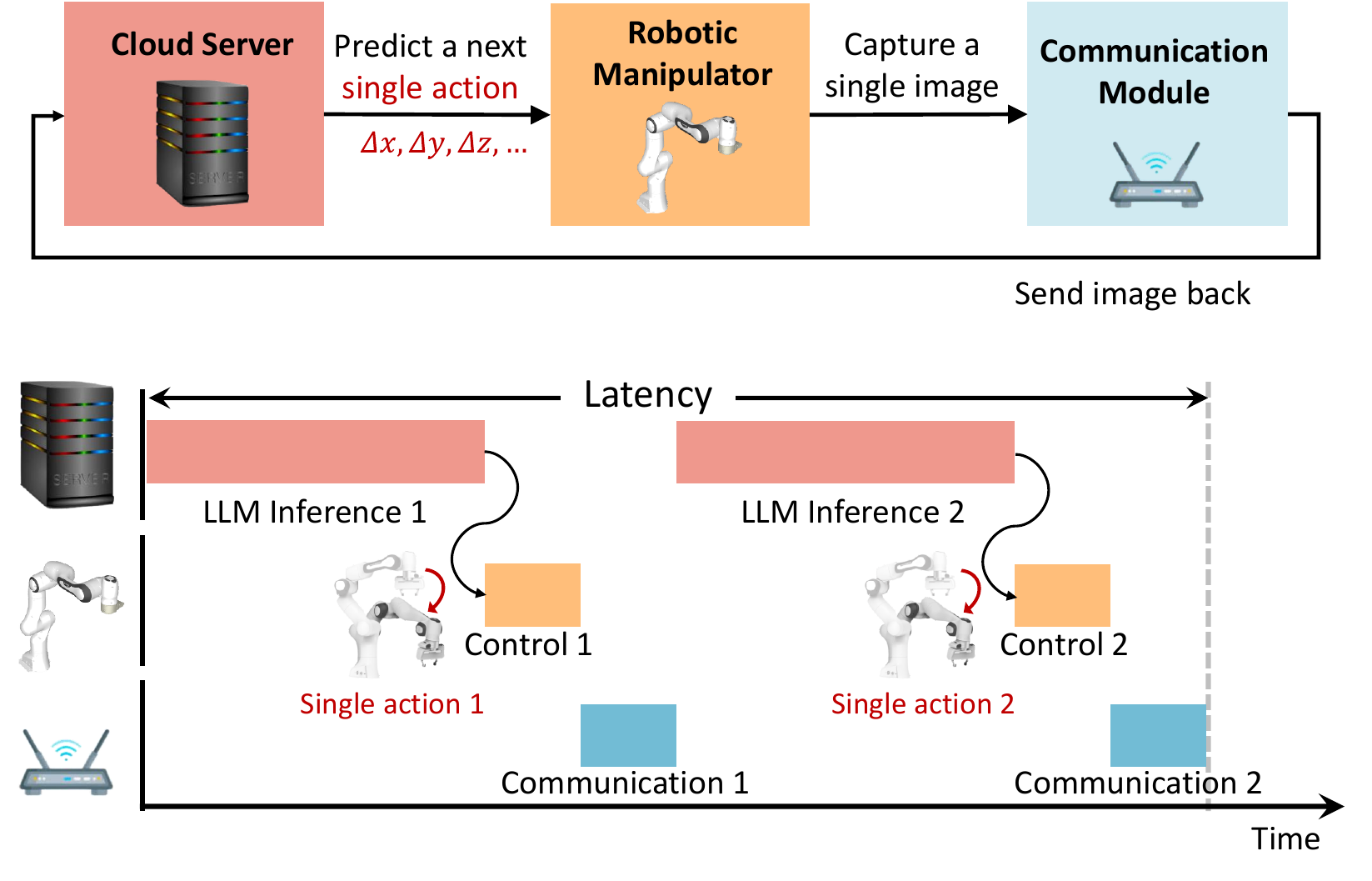}
  \label{fig:oldpipe}
}
% \hspace{2pt}
% \subfloat[Trajectory optimized using $\SE$ pose representation.]
% {
%   \includegraphics[trim=0 0 0 0, clip, width=0.30\columnwidth]{fig/spheres_2.pdf}
%   \label{fig:se3_sphere}
% }
\hspace{2pt}
\subfloat[Proposed continuous execution pipeline of embodied AI systems, where the model predicts near future trajectory and pipelines communication latency with robot execution latency.]
{
  \includegraphics[trim=0 0 0 0, clip, width=\columnwidth]{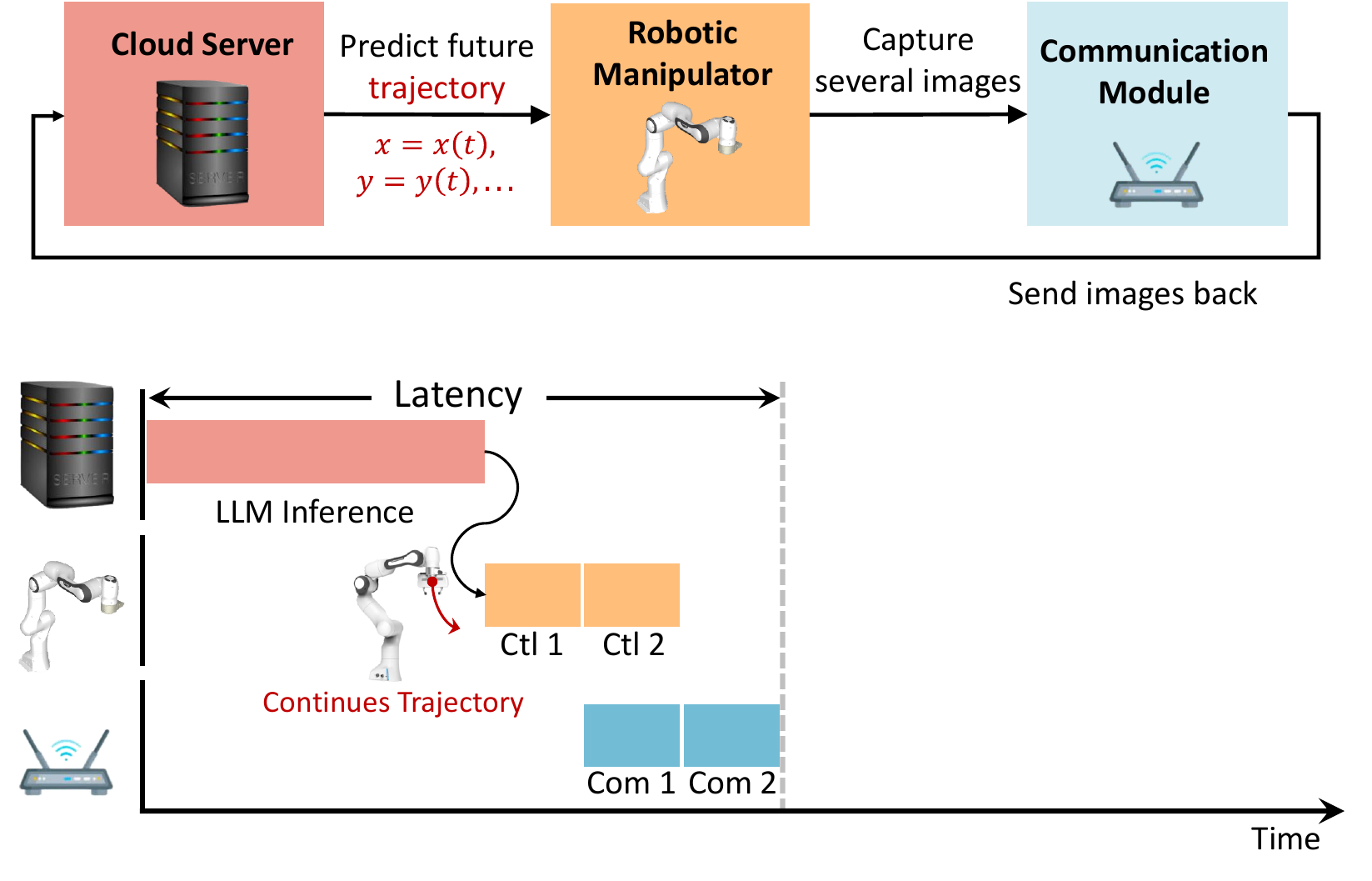}
  \label{fig:newpipe}
}
\caption{Existing embodied AI systems pipeline and \textsc{Corki}\xspace pipeline.}
\label{fig:sphere}
\end{figure}

The current execution model is time-consuming due to two primary reasons. First, the sequential nature of each stage significantly contributes to the overall latency. Since most robots depend on high-end servers for LLM inference, the latency associated with the embodied AI systems is the cumulative effect of three distinct stages: LLM inference latency, robot action execution latency, and data communication latency. The sum of these latencies for each frame can add up to several hundred milliseconds. Second, all three stages have to happen for every frame, further hurting the real-time constraints. We show this pipeline in Fig.~\ref{fig:oldpipe}. %Fig.~\ref{fig:oldpipe} illustrates the execution model of existing embodied AI systems. 

\paragraph{Idea.} Today's embodied AI pipeline is designed purely based on the convenience of algorithm designers, as executing frame by frame sequentially is a traditional method in video processing algorithms. Yet, it does not follow the design methodology in robotic domain and violates a basic principle of robotic software design that is widely adopted in robotic domain. The front-end, responsible for perception and planning, does not inherently require real-time performance. In contrast, the back-end, which includes robot control algorithms, must operate in real-time. The front-end and back-end can be aligned through a common intermediate representation: the robot's movement trajectory. Critically, the unbalanced frequency requirements in the robotic software stack motivate us to decouple LLM inference, robotic control, and data communication. After decoupling, we can reduce the front-end LLM inference rate, pipelining three stages and accelerating robotic control algorithms to achieve real-time performance in embodied AI applications. 

\paragraph{Design.} In this paper, we fundamentally change the execution pipeline of existing embodied AI systems to reduce the end-to-end latency. Firstly, at the algorithm level, we depart from the conventional approach of predicting robot movement in the next frame discretely. Instead, we propose a novel embodied AI algorithm framework that is able to predict the trajectory of the robot for the near future. Unlike methods that focus on predicting only the immediate subsequent step, our algorithm accurately forecasts actions for multiple future steps. Thus, we significantly reduce the inference frequency of LLMs, saving both latency and energy.

Secondly, to accelerate the control process, we devise an accelerator capable of translating the trajectories predicted by LLMs into seamless and real-time control signals. The accelerator we design is tailored for task space computed torque control, with a customized data-flow accelerator, customized circuits, and dedicated on-chip buffer design. The most crucial architectural insight we find is that robotic control computations are performed at high frequencies, yet each control signal's actual degree of change remains relatively low. This allows us to implement an application-specific approximate computing strategy. By analyzing the impact of individual joint movements on the control parameters, we can dynamically determine when to recompute these parameters and when to reuse previously computed values. This approximation significantly reduces computational overhead without compromising control accuracy.

Finally, at the system level, we streamline the transmission of newly captured frames to the server concurrently with the robot execution process. This approach effectively hides communication latency beneath the robot execution latency, resulting in a further reduction of the end-to-end latency. We illustrate our idea with Fig.~\ref{fig:newpipe}.

\begin{sloppypar}
\paragraph{Results.} We use a state-of-the-art embodied AI system, RoboFlamingo~\cite{livision}, as our baseline.
% \textsc{Corki}\xspace on top of it. We show that \textsc{Corki}\xspace is able to achieve $9.1\times$ speed up.
\textsc{Corki}\xspace largely reduces LLM inference frequency by up to $5.1\times$, resulting in up to $5.9\times$ speed up. The success rate improvement can be up to 13.9\%.
The maximum success rate improvement is 17.3\% higher than the baseline. The contribution of this paper is summarized as follows: 
\end{sloppypar}
\begin{itemize}
    \item We observe that the existing embodied AI pipeline can not satisfy real-time constraints because currently the pipeline design is vision-centric, operating on a frame-by-frame basis, which results in high frame latencies.
    \item We propose a new embodied AI algorithm framework from a robotic-centric angle to control robots by predicting future trajectories instead of the discrete movement of every frame, combined with a classic (non-LLM) high frequency controller. 
    \item We design a new execution pipeline based on our proposed framework to hide communication latency between the robot body and the server.
    \item We design an accelerator to smoothly transform the trajectory predicted by our models into robotic control signals in real-time.
    \item We demonstrate \textsc{Corki}\xspace with an efficient implementation of the proposed embodied AI system. We show that \textsc{Corki}\xspace is able to significantly reduce the end-to-end latency without sacrificing accuracy. 

\end{itemize}

We organize our paper as follows. Sec.~\ref{sec:baseline} introduces basic embodied AI system pipeline and motivates our paper. Sec.~\ref{sec:algo} introduces a new embodied AI algorithm framework that is able to predict the continuous near-future trajectory of robots. Sec.~\ref{sec:arch} describes the proposed hardware accelerator for controlling robots given predicted trajectory and system pipeline design. Sec.~\ref{sec:setup} discusses the experimental methodology, followed by the evaluation results in Sec.~\ref{sec:eval}. We discuss the related work in Sec.~\ref{sec:related} and conclude our paper in Sec.~\ref{sec:concl}.
\section{Background and Motivation}
\label{sec:baseline}

We introduce the background of embodied AI systems (Sec.~\ref{sec:baseline:emai}). We show that the execution pipeline of embodied AI systems is significantly different from the previous utilization of LLMs and results in high end-to-end frame latency (Sec.~\ref{sec:baseline:moti}). 

\subsection{Embodied AI System}
\label{sec:baseline:emai}
For manipulation tasks, traditional robots typically depend on rule-based algorithms for decision-making and task planning, confining their utility to simple and predetermined scenarios. In contrast, the success of Large Language Models (LLMs) has spurred efforts to equip robots with advanced reasoning and long-term planning capabilities. Such success boosts the emergence of applications that use LLMs for robot control, which has demonstrated notable advancements, particularly in enhancing the success rates of robots performing complex tasks in dynamic scenarios~\cite{livision,wake2023gpt,hu2023toward,firoozi2023foundation}.

Embodied AI systems represent a category of systems that leverage the reasoning abilities of Large Language Models (LLMs) to guide robots in accomplishing complex real-world tasks, including but not limited to housekeeping and industrial manufacturing, with the goal of reducing human efforts. Typically, these systems comprise two integral components: a high-end server equipped with GPUs for LLM inference and a robot body responsible for executing and interacting with the physical environment.

Embodied AI systems commonly employ a multi-modality Large Language Model ~\cite{lyu2023macaw,zhao2023bubogpt,mu2024embodiedgpt,huang2023chatgpt} as the planning module. This LLM seamlessly integrates language instruction inputs, such as "put the blue mug on the table and bring me the red one," with traditional sensor inputs in the robotic pipeline, including continuous videos, IMU signals, and point clouds~\cite{everett1995sensors,li2019common,santaera2015low}. The LLM inference will generate the next actions for the robot body to perform based on current and recent observations along with the instructions. 

Recently, embodied AI systems have demonstrated substantial potential to replace humans in various tasks. Google's robotic transformer~\cite{RT1,RT2} has achieved an impressive success rate of over 75.0\% on tasks including "pick up objects", "open drawers", and "place objects into designated places" within domestic environments. RoboFlamingo~\cite{livision}, a recently proposed embodied AI framework, further elevates the success rate of a single task to over 89.5\%. 

\subsection{Execution Pipeline and Performance Bottleneck}
\label{sec:baseline:moti}
We use RoboFlamingo as an example of embodied AI systems to illustrate existing system pipelines. RoboFlamingo utilizes a vision-language model (VLM) to control a Franka Emika Panda robot arm with a parallel gripper~\cite{gaz2019dynamic}, which in total has seven degrees of freedom. RoboFlamingo takes two forms of input: a language instruction and a video sequence containing 12 images. The model will predict the action of the robot arm's end-effector within the next step. Equ.~\ref{eq:llm_inf} describes the LLM inference process at frame $t$, where $F_{t}$ represents a single frame within a video sequence and $i$ denotes the language instruction. $\Delta x, \Delta y, \Delta z$ are the three-dimensional position change, $\Delta\alpha, \Delta\beta, \Delta\gamma$ are the three-dimensional rotation change, and $g$ is the one-dimensional gripper status, which can be either open or closed. %0 represents close the gripper and 1 represents open the gripper. 

After the model predicts the action, the robot arm will perform the action, moving itself to a new position. The control process on the robot translates the movement information of the end-effector to the actual torque of each motor placed on the joints of the robot arm. A camera on the robot, usually placed on the gripper, will capture a new frame $F_{t+1}$ and send it back to the model to update the input frames. The next inference will happen on $(F_{t-10}, F_{t-9}, ..., F_{t+1},i)$. 

\begin{figure}[t]
%\vspace{-10pt}
\centering
\subfloat[\small{Per-frame latency breakdown.}]
{
  \includegraphics[trim=0 0 0 0, clip, width=0.5\columnwidth]{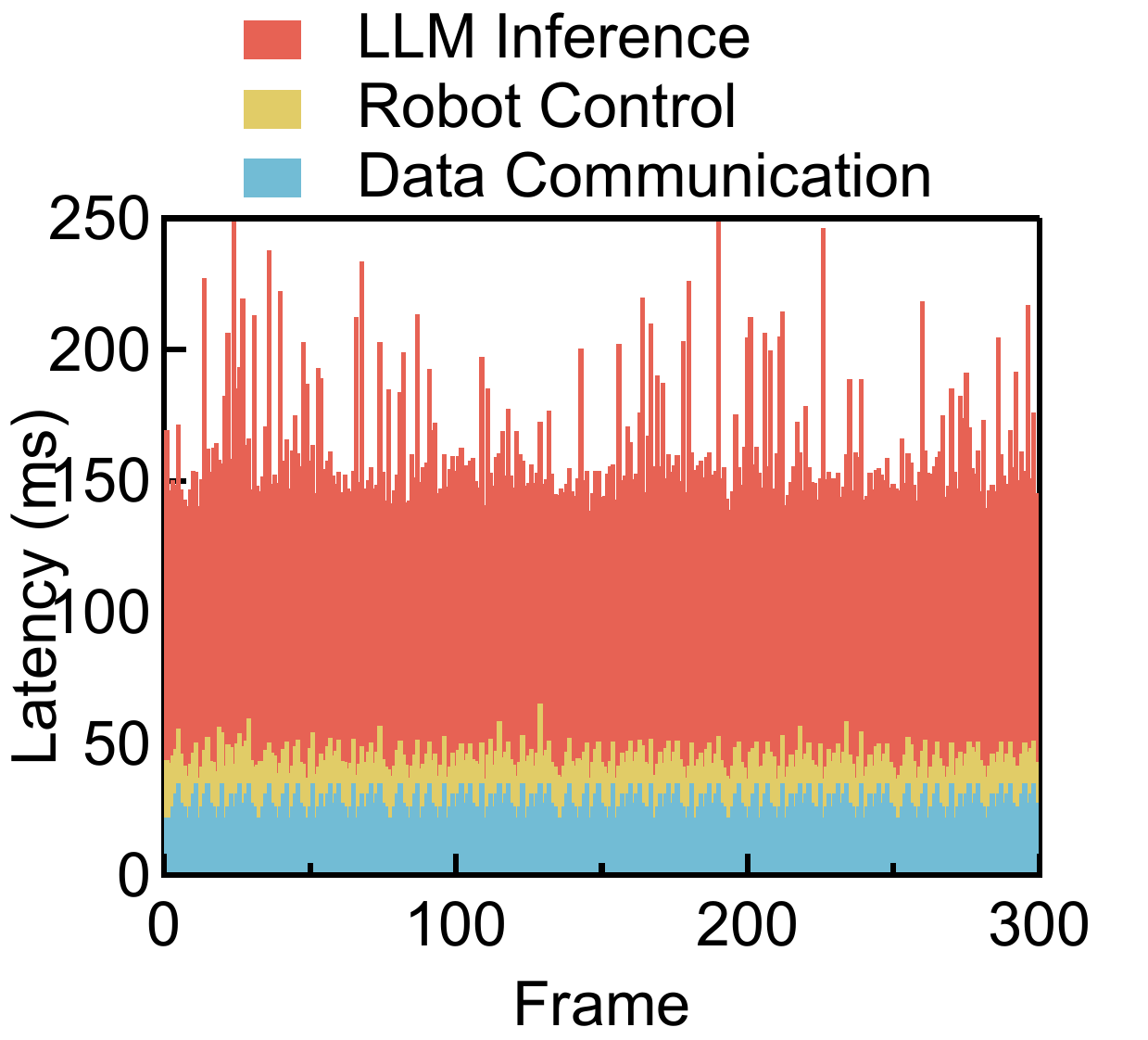}
  \label{fig:moti_latency}
}
%\hspace{3pt}
\subfloat[\small{Per-frame energy breakdown.}]
{
  \includegraphics[trim=0 0 0 0, clip, width=0.5\columnwidth]{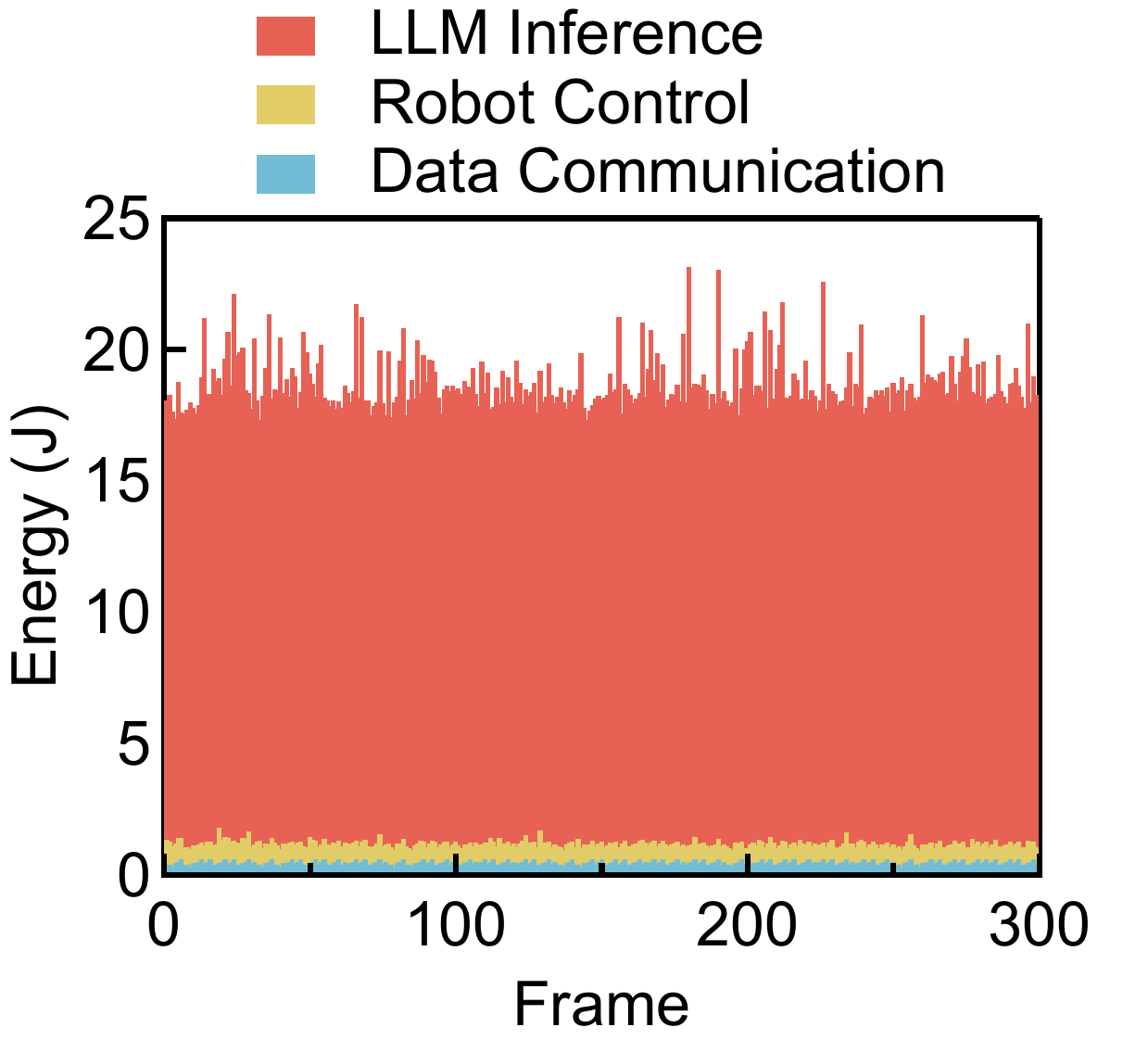}
  \label{fig:moti_energy}
}
\caption{Latency and energy breakdown of RoboFlamingo.}
\label{fig:moti}
\vspace{-10pt}
\end{figure}

\begin{align}
    % (\Delta x_t,\Delta y_t,\Delta z_t,\Delta \alpha_t ,\Delta \beta_t,\Delta \gamma_t,g) = LLM(F_{t-11},F_{t-10},...,F_{t},i)
    % (\Delta p, \Delta r, g) = LLM(F_{t-11},F_{t-10},...,F_{t},i)
    (\Delta x,\Delta y,\Delta z,\Delta\alpha, \Delta\beta, \Delta\gamma, g) = LLM(F_{t-11},F_{t-10},...,F_{t},i)
    \label{eq:llm_inf}
\end{align}

Specifically, we analyze and characterize the execution pipeline of RoboFlamingo by breaking down the execution latency with the results presented in Fig.~\ref{fig:moti}. To get the results, we run LLM inference on a Nvidia V100 GPU, the robot control on a Franka Emika Panda robot arm, which is equipped with an Intel Core i7-6770HQ CPU and use the CPU to process control algorithms, and gather the communication data using a Wi-Fi module. The latency and power data are first measured, the energy consumption is then calculated using the equation $E = \int_{0}^{T} P(t) dt$.

Fig.~\ref{fig:moti_latency} shows that even with a relatively small LLM (3 billion parameters) and a high-end GPU, the end-to-end frame latency of the embodied AI system can reach up to 249.4 ms, which directly contributes to a very low frame rate that does not satisfy real-time constraints. Among all three stages, LLM inference takes 72.7\% of the execution time, robot control takes 9.9\%, and data communication takes 17.4\%. Fig.~\ref{fig:moti_energy} shows the energy breakdown. LLM inference still dominates with 95.8\% of the total energy, while robot control and data communication account for only 4.2\%. Notice that the latency spent on control is low in the baseline system since the control frequency is set to match the front-end frame rate of 30 Hz. However, in real robotic systems, control usually has a much higher rate. Our characterization suggests that for each frame, to get a smooth trajectory, corresponding control latency can add up to 13.9\% of the total latency. 
%Robot control comes next with 1\% and data communication only takes \fixme{2\%}. 

\paragraph{Bottleneck Analysis.} Detailed characterization data suggests that the reasons for the slow execution of embodied AI robots are mainly twofold. First, the frame-by-frame sequential execution pipeline forces every action of the robot to undergo three stages: LLM inference, robot control and communication, and the latencies accumulate. Second, LLM inference, even with high-end GPU acceleration, is still extremely slow. 

The motivation for dedicated accelerators for control is clear. One of the key contributions of this work is reducing LLM inference frequency. However, our characterization shows that even if LLM inference latency were reduced to zero, the control frequency would still only reach 22.1 Hz, falling short of the real-time requirement (at least 30 Hz, 100 Hz preferable), while higher control rate allows the robot to better follow the predicted trajectory, increasing the safety of the control. Furthermore, control operations account for 39.7\% of the total latency, with the rest latency spent on communication. The above reasons motivate the accelerator design. Note that we use the Intel i7-6770HQ CPU only because it is the onboard processor of our robot. We also tried to process the control algorithm with an Intel Core i7-13700 CPU and the corresponding frequency still can not meet real-time requirements.

From the perspective of a robotic system designer, the planning module does not need to match the high frequency of the control module. Trajectory is usually used as a bridge to eliminate the frequency mismatch. We apply the same principle. 
\section{\textsc{Corki}\xspace Algorithm Framework}
\label{sec:algo}
We introduce \textsc{Corki}\xspace algorithm in this part. The key insight of our algorithmic innovation is to change per-frame robot action prediction (Sec.~\ref{sec:algo:base}) into robot trajectory prediction (Sec.~\ref{sec:algo:traj}). We further optimize the algorithm framework with an adaptive trajectory length selection (Sec.~\ref{sec:algo:adap}), which also provides accuracy and performance trade-off.

%Further, to improve performance and accuracy trade-off, we propose to select the length of the trajectory length adaptively (Sec.~\ref{sec:algo:adap}). 

\subsection{Baseline Embodied AI Algorithms}
\label{sec:algo:base}

RoboFlamingo is comprised of two main components: a vision language model (VLM)~\cite{alayrac2022flamingo} and an LSTM network~\cite{hochreiter1997long} named policy head. At every time step $t$,  the VLM takes visual observations $F_t$ and a language instruction $i$ as input and outputs vision-language tokens $X_t$. The robot actions $a_t$ are generated through the policy head using given $X_t$~\cite{livision}.

%Flamingo \cite{alayrac2022flamingo} for the VLM $f_{\theta}$ and LSTM \cite{hochreiter1997long} for the policy head $p_{\theta}$. During each time step $t$,  the VLM takes visual observations $F_t$ and a language instruction $i$ as input, and outputs vision-language tokens $X_t$. These tokens are then mapped to the robot action $a_t$ by the policy head, enabling the robot to follow the given instruction.$i$\cite{li2023vision}.

\begin{figure}[t]
    \centering
    \includegraphics[width=0.9\columnwidth]{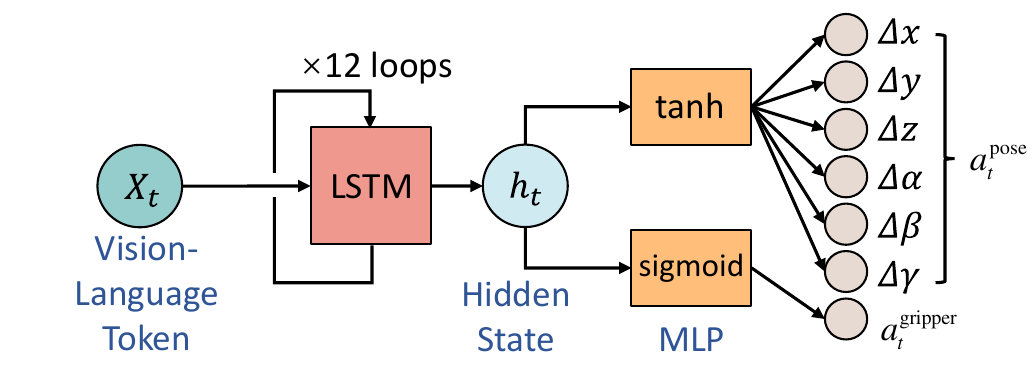}
    \caption{RoboFlamingo policy head. The vision-language token is from the LLM inference. The outputs are seven-dimensional variables representing the robot's movements in the next time step.}
    \label{fig:lstm1}
\end{figure}

We elaborate on the action generation process in Fig.~\ref{fig:lstm1}. At each time step $t$, the policy head takes the visual-language tokens $X_t$ generated by the LLM as input and goes through an LSTM network. The hidden state $h_t$ is then mapped to the 7-DoF action through two MLP heads, as shown in Equ.~\ref{equ:act}:

%$\tilde { X } _ { t } = \operatorname { MaxPooling } ( X _ { t } )$ as input. After passing through the LSTM network: $ h _ { t } = \operatorname { LSTM } ( \tilde { X } _ { t }, h _ { t - 1 } )$, the hidden state is mapped to the 7-DoF action through two MLP heads: 

\begin{equation}
\label{equ:act}
     a _ { t } ^ { \text { pose } }, a _ { t } ^ { \text { gripper } }= \operatorname {MLP} ( h_t ).
\end{equation}

The training loss thus contains two parts, as illustrated in Equ.~\ref{eq:loss_base}, the pose estimation is supervised using mean squared error (MSE) loss, while the gripper status is supervised using binary cross-entropy (BCE) loss. The weight $\lambda$ is used to balance the two parts.

%Fig.~\ref{fig:lstm1} indicates the details of the policy head. In the prediction of each time step $t$, the policy head takes the visual-language tokens $\tilde { X } _ { t } = \operatorname { MaxPooling } ( X _ { t } )$ as input. After passing through the LSTM network:$ h _ { t } = \operatorname { LSTM } ( \tilde { X } _ { t } , h _ { t - 1 } )$, the hidden state is mapped to the 7-DoF action through two MLP heads: $ a _ { t } ^ { \text { pose } },a _ { t } ^ { \text { gripper } }= \operatorname {MLP} ( h_t )$. Besides, there are some settings to ensure that training and inference work the same way. During the training process, each training video sequence contains 12 corresponding observations and action labels. With the continuous input of images, the LSTM network will loop out the corresponding action of each frame, then utilize maximum likelihood imitation learning objectives to fine-tune the backbone. The loss consisting of mean squared error (MSE) loss and binary cross-entropy (BCE) loss as \eqref{eq:loss_base} , $\lambda$ corresponds to the weight of gripper loss\cite{li2023vision}.

\begin{align}
    \ell = \sum _ { t } \operatorname { MSE } ( a _ { t } ^ { \text { pose } } , \hat { a } _ { t } ^ { \text { pose } } ) + \lambda  \operatorname { BCE } ( a _ { t } ^ { \text { gripper } } , \hat { a } _ { t } ^ { \text { gripper } } ) \label{eq:loss_base}
\end{align}

During inference, the policy head maintains a queue of length 12. If the queue is not full, the policy head will predict the action $a_t^{\text {pose}}$, $a_t^{\text {gripper}}$ and update the hidden state $h_t$ for the next step prediction; once the queue reaches its maximum capacity, the earliest tokens that entered the queue will be replaced by the latest ones, then, consistent with the training process, the action of the current step $a_t$ is given based on the last 12 vision-language tokens $X_{t-11} \thicksim X_{t}$.

\subsection{Basic \textsc{Corki}\xspace Algorithm}
\label{sec:algo:traj}
We think the fundamental design principle of current embodied AI algorithms is to better supervise the output of every frame. However, the frame-by-frame supervision violates the philosophy of the robotic system. We thus introduce to predict trajectory instead, describe the corresponding training modifications, and further improve our design through an adaptive trajectory length decision during runtime. 

%In this section, we will introduce the details of our algorithm. By changing frame-by-frame prediction to trajectory prediction, the algorithm can predict a continuous trajectory, thus reducing the number of times the LLM is executed. Here are three parts. First, we will introduce the trajectory prediction algorithm and its loss supervision mode.Second, we will introduce the variation we use and the masked policy head algorithm, and finally, we will introduce the mask embedding training mode and conduct experimental analysis.

\paragraph{Trajectory Prediction.}
We predict the continuous trajectory of the nearest future instead of discrete actions. We use a cubic function to fit the motion trajectory of robotic arms. For all the seven variables we need to predict, we output a trajectory for each one of the first six variables ($r_{x}(t),r_{y}(t),r_{z}(t),r_{\alpha}(t) ,r_{\beta}(t),r_{\gamma}(t)$), the gripper $g$ is still a binary value. Using $r_{x}(t)$ as an example, the model output will be shown as Equ.~\ref{eq:episode}, where $t$ represents time. 

%We convert the prediction of a future frame of discrete action into a continuous trajectory of several frames. The cubic function is chosen to fit the robot motion because its flexibility, continuity and interpolation properties can well meet the trajectory fitting requirements.
%Given the action space  $a_{t}^{pose} = (\Delta x_t,\Delta y_t,\Delta z_t,\Delta \alpha_t ,\Delta \beta_t,\Delta \gamma_t)$, \fixme{inconsistent with the previous symbol in Equ.~\ref{eq:llm_inf}?} we fit each dimension with a trajectory. Suppose that our predicted trajectories contain N previously discrete actions , then we have variable $ k = 0,1,...,N - 1$. As shown in \eqref{eq:episode}, the policy head will predict the trajectory parameters $a,b,c,d$, we then input the time step variable $k$ to calculate the action $\Delta X_k$to be executed.

\begin{align}
r_{x}(t) = a t ^ { 3 } + b t ^ { 2 } + c t + d \label{eq:episode}
\end{align}

We employ the cubic function as the trajectory function for two key reasons. First, in robot trajectory planning,  the cubic function inherently captures changes in velocity and acceleration since the derivative and second derivative of the cubic function are continuous, effectively modelling the dynamics of real-world motion. Second, they help mitigate overfitting and enhance robustness to noise.

\paragraph{Loss Design.}
% After we change the model output, there are two ways of designing loss. The first one is directly supervising $a,b,c,d$. The second one is to supervise the trajectory with the ground truth. We go for the second one for two reasons. The first reason is that usually, no dataset provides the $a,b,c,d$ ground truth, which means we need to extract it from the trajectory ground truth first, which accumulates errors. Second, these parameters vary significantly and are not conducive to the learning of the neural network.
After we change the model output, there are two ways of designing loss. The first one directly supervises $a,b,c,d$. The second one is to supervise the trajectory with the ground truth. We go for the second one for two reasons. The first reason is that almost no dataset provides the $a,b,c,d$ ground truth, so we need to extract it from the trajectory ground truth first, accumulating errors. Second, these parameters vary significantly and are not conducive to the neural network's learning process.

%When designing for loss, we do not directly monitor the parameters$a,b,c,d$ for two reasons.1.In our experiment, the cubic function is too flexible, resulting in trajectory parameter labels obtained from the trajectory fitting algorithm differing from the expected trajectories. 2.the trajectory parameters change too much, which may not be conducive to the learning of the neural network. Therefore, we have chosen to supervise the points on the trajectory.\fixme{Need a figure here to show the gt have problems?}

Using variable $r_{x}(t)$ as an example. We supervise the trajectory action $T_x$ in the training set and our predicted trajectory $r_x$ using the MSE shown in Equ.~\ref{eq:loss_x}. Then, we update our trajectory parameters through backpropagation. In this way, we no longer need to get discrete actions with 30 Hz first and can directly monitor the trajectory to obtain a more capable model.

%We multiply these parameters with the time step variable $k$ to obtain the predicted action $\Delta X_k$, and supervise it using equation \eqref{eq:loss_x}.

% \begin{gather}
% \Delta \hat X _ { t + k } = \sum _ { i = 0 } ^ { k } \Delta \hat x _ {t + i }\label{eq:sum} \\ 
% \ell_{x} = \sum _ {j = 0} ^ {k}\operatorname { MSE } ( \Delta X_ { t + j }  , \Delta \hat X_{ t + j} )  \label{eq:loss_x} 
% \end{gather}

\begin{gather}
\ell_{x} = \sum _ {j = 0} ^ {k}\operatorname { MSE } ( r_{x}(j) , T_{x}(j))  \label{eq:loss_x} 
\end{gather}

\begin{figure}[t]
    \centering
    \includegraphics[width=0.9\columnwidth]{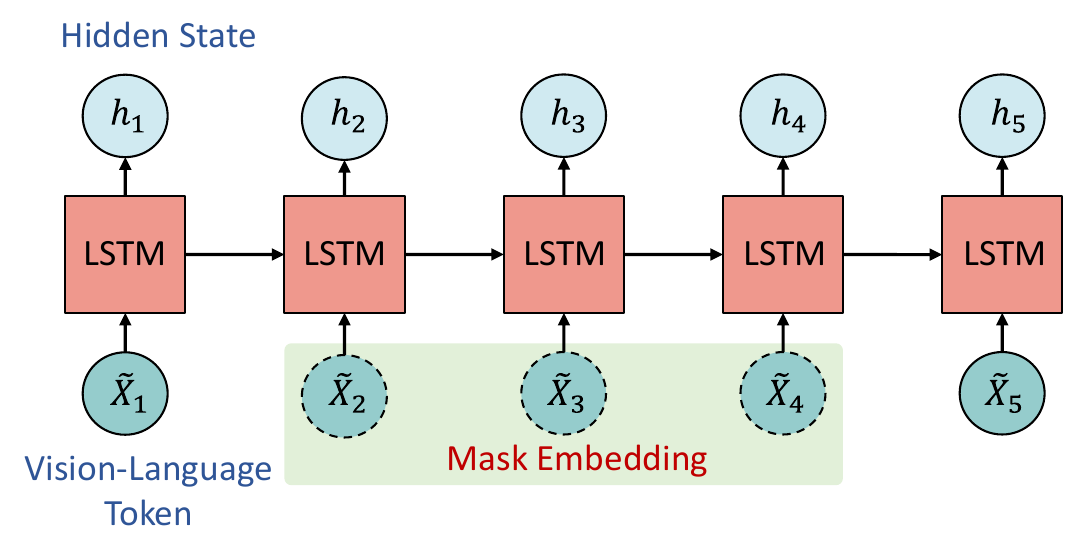}
    \caption{Masked policy head. The tokens in the dotted line are not generated through a LLM but instead a mask embedding.}
    \label{fig:lstm2}
\end{figure}

Because of our design, the robotic control and vision inputs are decomposed, leading to less information captured by the robots. To mimic this process during training, we intentionally insert fewer images. As shown in Fig.~\ref{fig:lstm2}, vision-language tokens from $t=2$ to $t=4$ are shed by a mask embedding, similar to existing works such as BERT~\cite{kenton2019bert}. 
\label{sec:mask embedding}

\subsection{Optimizing \textsc{Corki}\xspace Algorithm}
\label{sec:algo:adap}
In the basic \textsc{Corki}\xspace algorithm, the length of the trajectory is fixed all the time. Suppose the prediction interval is set to be 165 ms. No shorter or longer trajectory can be taken. However, one of the most significant characteristics of robotic applications is that they usually encounter sudden environmental changes. 

\paragraph{Early Termination. } We thus provide flexibility in the length of the trajectory that is taken. The prediction length will be used as an upper bound of the length of the actual taken trajectory, and early termination is allowed. Again, assuming the prediction interval is 165 ms, the actual trajectory can be from 33 ms to 165 ms, with a stride of $\Delta t$ (which is 33 ms, assuming the camera sensor works in a 30 Hz frequency). After the robot's early termination, the model will predict another trajectory for the 165 ms. 

Early termination gives us some flexibility, but it may not be enough. The reason is that the accuracy is higher when the actual trajectory length is consistent between training and inference. For example, in training, we predict 5 frames, later in testing, although generalizable, fixing the five-step termination shows better accuracy than other step terminations (like 3 or 4). If the actual trajectory length is 66 ms in training, the same length should be taken during inference. Suppose the user wants to change the actual trajectory length. In that case, the only way is to train two models, one for 66 ms and one for 99 ms, and switch during inference, which is unsurprisingly inconvenient for almost all robotic applications.  

%\paragraph{Take N Predict M.}
%In order to compare different experimental configurations, we set several variations of our algorithm. Two key configurations that need clarification are the predict steps $N$, and the take steps $M$. The predict steps $N$ refers to the action steps predicted by our model, as shown in the figure Fig.~\ref{fig:multistep} above. 

%\begin{figure}[htbp]
%    \centering
%    \includegraphics[width=\columnwidth]{figs/multistep.pdf}
%    \caption{example of predict N take M}
%    \label{fig:multistep}
%\end{figure}

%On the other hand, take steps $M$ is the number of steps we carried out in the simulation environment. It is important to note that in our setup, each take step does not require intermediate frames, indicating that our configurations involve open-loop control.Based on the above experimental results, In addition to balancing energy consumption and frequency, we have developed an adaptive algorithm detailed in Sec.~\ref{sec:algo:adap}.

%\paragraph{Masked Policy head.}
%In order to make the policy head aware of the time step change, in the execution step, we mask the unseen time steps observation with masked tokens in the open-loop control, and the output corresponding to these masked tokens will still be monitored as Fig.~\ref{fig:lstm2}. Similar to the mask mode in BERT\cite{}, the masked token parameters will be updated during the back propagation. This approach allows us to improve the model's performance while keeping latency unchanged significantly.

\paragraph{Adaptive Trajectory Length.} Our method is to increase flexibility by allowing adaptive trajectory length with an empirical method. Our insight comes from the curvature of the trajectory. When the curvature is low, the action does not change significantly, suggesting a longer trajectory is acceptable. However, when the curvature is high, the usual circumstance is that the robot is encountering sudden change, where a shorter trajectory is better. 

\begin{figure}[t]
    \centering
    \includegraphics[width=\columnwidth]{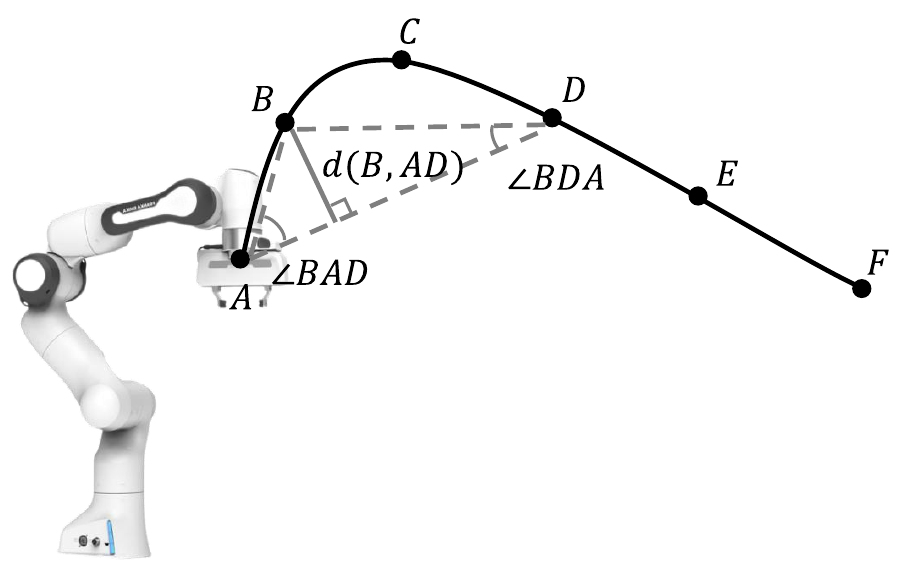}
    \caption{Waypoints extraction and identification algorithm. The first waypoint with huge movement will be identified and taken as the endpoint of the trajectory.}
    \label{fig:waypoints}
    % \vspace{-25pt}
\end{figure}

\paragraph{Waypoints Extraction.} We identify the adaptive trajectory length using a concept called waypoints. For example, for a given trajectory spanning 165 ms, a waypoint is defined as a point on the trajectory every 33 ms or each time step. In Fig.~\ref{fig:waypoints}, point $A$ is the starting point, points $B$ to $F$ are the waypoints, and point $F$ is the endpoint. Waypoint identification aims to find a waypoint where the robot's movement is significant. In our case, significant movements are identified as high curvature or changes in the gripper state. The reason is that high curvature suggests a rapid change in the robot's intended motion, which is often necessary to react to new obstacles, adjusted goals, or other unexpected events. By terminating the current trajectory at a point of high curvature, we allow the system to re-evaluate the situation with updated sensor information and generate a new trajectory, enhancing the robot's ability to adapt to dynamic environments.

%\origin{which position's curvature is too high} \yiyang{which position's with varying movements, including high curvature and gripper state
%changes}, so that the actual taken trajectory should end at this waypoint, instead of predicted F. 

\paragraph{Waypoints Identification.} We check each waypoint from $B$ to $F$ and compare two metrics to identify potential waypoints with high curvature. Given the example in Fig.~\ref{fig:waypoints}, the current waypoint undergoes checking is $D$. For every point in the interval of $[B,D)$, we compare two metrics with corresponding thresholds. The first is the $\angle{BAD}$ and $\angle{BDA}$ with a threshold of 90 degrees. The second one is the distance between point $B$ to line $AD$, or $d(B, AD)$ with a threshold $d$. If any threshold is violated, we consider the curvature at a point between $C$ and $D$ to be high, and thus, the trajectory should end at $D$ instead of the predicted point $F$. The length of the trajectory depends on the endpoint we get. 

To find potential waypoints with gripper state changes, we compare the state of the gripper at the current waypoint and the next waypoint. If the gripper states of these two waypoints are different, the current waypoint will be identified as one with significant movement. 

%\yiyang{For waypoints involving a change in the gripper state, we will designate the position where the gripper closes and the position immediately before the gripper closes as waypoints. For example in Fig.~\ref{fig:waypoints}, current
%waypoint undergo checking is $D$.if point $E$'s gripper state $g(E) = 1$, which means the gripper is closed, then points $D$ and $E$ are considered waypoints to help the model capture changes in these critical positions.}

%\paragraph{Adaptive trajectory algorithm.} \label{sec:algo:adap}
%The goal of this paragraph is to introduce our adaptive algorithm. Based on the experimental results of Table \ref{D-D}, we found a trade-off between task success rate and model frequency. Driven by the need to optimize success rate and frequency, We propose our adaptive trajectory algorithm. First, we specify the waypoint extraction rules and then show how to use this simple key point extraction algorithm for two-stage training to help improve the model's performance.

%\paragraph{Waypoints extraction algorithm}
%The algorithm's main objective is to identify waypoints in the trajectory. As shown in algorithm \ref{algo:waypoint} and Fig.~\ref{fig:waypoints}.

%The concept behind waypoint extraction is based on the insight that areas with more pronounced curvature in a robot's motion trajectory should be observed more frequently. On the other hand, areas with less curvature can be moved through more quickly to reduce the number of observations.

\begin{algorithm}[htbp]   
    \caption{Adaptive trajectory length.}
    \label{algo:waypoint}
        \renewcommand{\algorithmicrequire}{\textbf{Input:}}
    \renewcommand{\algorithmicensure}{\textbf{Output:}}
    
\begin{algorithmic}[1]
        \REQUIRE Starting Point $A$, Trajectory $T$ \\
         Gripper states $G$ = {0,0,0,1,0}
        \ENSURE The earliest termination point  %%output 
\\ //Extracting waypoints at each time step.\\
$B,C,D,E,F$ = $E(A,T)$
\FOR{$P$ in the range[$B, F$)}
    \STATE $P_{n}\gets$ the next waypoint of $P$
    \IF{$G(P) \, \OR \, G(P_{n}) = 1 $  }
            \RETURN $P$
        \ENDIF
    \FOR{$p$ in the range (A,P]}
        \IF{$\angle(p,AP) \, \OR \, \angle(p,PA) > \frac{\pi}{2}$ || $D(p,AP) > d$}
            \RETURN $P$
        \ENDIF
    \ENDFOR
\ENDFOR
\RETURN $F$
\end{algorithmic}
\end{algorithm}

We explain the process in Algo.~\ref{algo:waypoint}. As the adaptive trajectory length is determined during runtime, the latency is thus sensitive. The algorithm we propose is effective and with low latency. In most cases, the total computational cost of Algo.~\ref{algo:waypoint} is less than 500 FLOPs.  

We provide users with an algorithm framework. Users can decide the length of trajectory prediction, whether early termination is needed, the level of early termination, and whether adaptive trajectory length is needed. 

\subsection{Close-loop Feature}
\label{sec:algo:close}
Till now, \textsc{Corki}\xspace lacks the ability to sense environmental changes during the trajectory prediction process. The algorithm generates trajectories, with no feedback until the next inference,  However, open-loop control can lead to potential error accumulation in robotic manipulation tasks, compared to close-loop control.

We thus introduce close-loop features. During the execution of each trajectory, we randomly send images back before the endpoint of the trajectory. These images are encoded using an encoder network ViT~\cite{dosovitskiy2020image}. The post-encoding close-loop features and tokens generated through the LLM are concatenated and used to predict the subsequent trajectory. These features will supervise the step size, and help \textsc{Corki}\xspace adjust the predictions in future iterations, accounting for potential changes in the environment.

\section{\textsc{Corki}\xspace Hardware and System Design}
\label{sec:arch}

%This section presents the \textsc{Corki}\xspace hardware. In the traditional implementation of task space computed torque control algorithms for robotic manipulators, although the algorithm design follows the principle of modularization, which facilitates programming for users as well as maintenance for developers, a large number of repetitive calculations are inevitably introduced. We address this problem by breaking the module hierarchy to enable data reuse (Sec.~\ref{sec:sec:break}). Besides, in high-frequency control environments, continuous real-time updating of parameters such as the Jacobian matrix and mass matrix ensures control accuracy but creates an unnecessary computational burden that affects the overall effectiveness of the system. We propose a dynamic updating mechanism for the parameters to solve the above problem (Sec.~\ref{sec:sec:dyna}).

This section introduces the \textsc{Corki}\xspace hardware. We accelerate the control process to achieve real-time performance. The input of the control module is the trajectory predicted by the \textsc{Corki}\xspace algorithm, and the output is torque signals that will be used on the motors in each joint of the robots. The control framework we build our accelerator on is widely used task space computed torque control (TS-CTC)~\cite{murray2017mathematical}. 
While the techniques in this section are broadly applicable beyond LLM settings, our accelerator is specifically designed to rapidly convert trajectories into control signals, which is crucial for real-time performance and central to our contributions.
We first elaborate on the control framework (Sec.~\ref{sec:sec:tsctc}), then analyze the bottleneck and propose the \textsc{Corki}\xspace accelerator (Sec.~\ref{sec:sec:break}). We further propose an effective approximation strategy to improve the control frequency (Sec.~\ref{sec:sec:dyna}). We finally describe the system pipeline (Sec.~\ref{sec:arch:pipe}).

%Given that the trajectories predicted by the \textsc{Corki}\xspace AI framework operate within the task space of the robotic manipulator, we utilize the task space computed torque control (TS-CTC) scheme ~\cite{}. TS-CTC computes the necessary joint torques according to the reference trajectories within the task space of the manipulator. We first describe the workflow of TS-CTC (Sec.~\ref{sec:sec:tsctc}), then analyze the issues present in the software implementation of TS-CTC, and  propose our hardware solution (Sec.~\ref{sec:sec:break}). Additionally, in high-frequency control, real-time updates of all model parameters in each control cycle impose a significant computational burden. To address this, we propose a dynamic parameter update strategy (Sec.~\ref{sec:sec:dyna}).

\subsection{Task Space Computed Torque Control}
\label{sec:sec:tsctc}
\begin{figure}[t]
    \centering
    \includegraphics[width=\columnwidth]{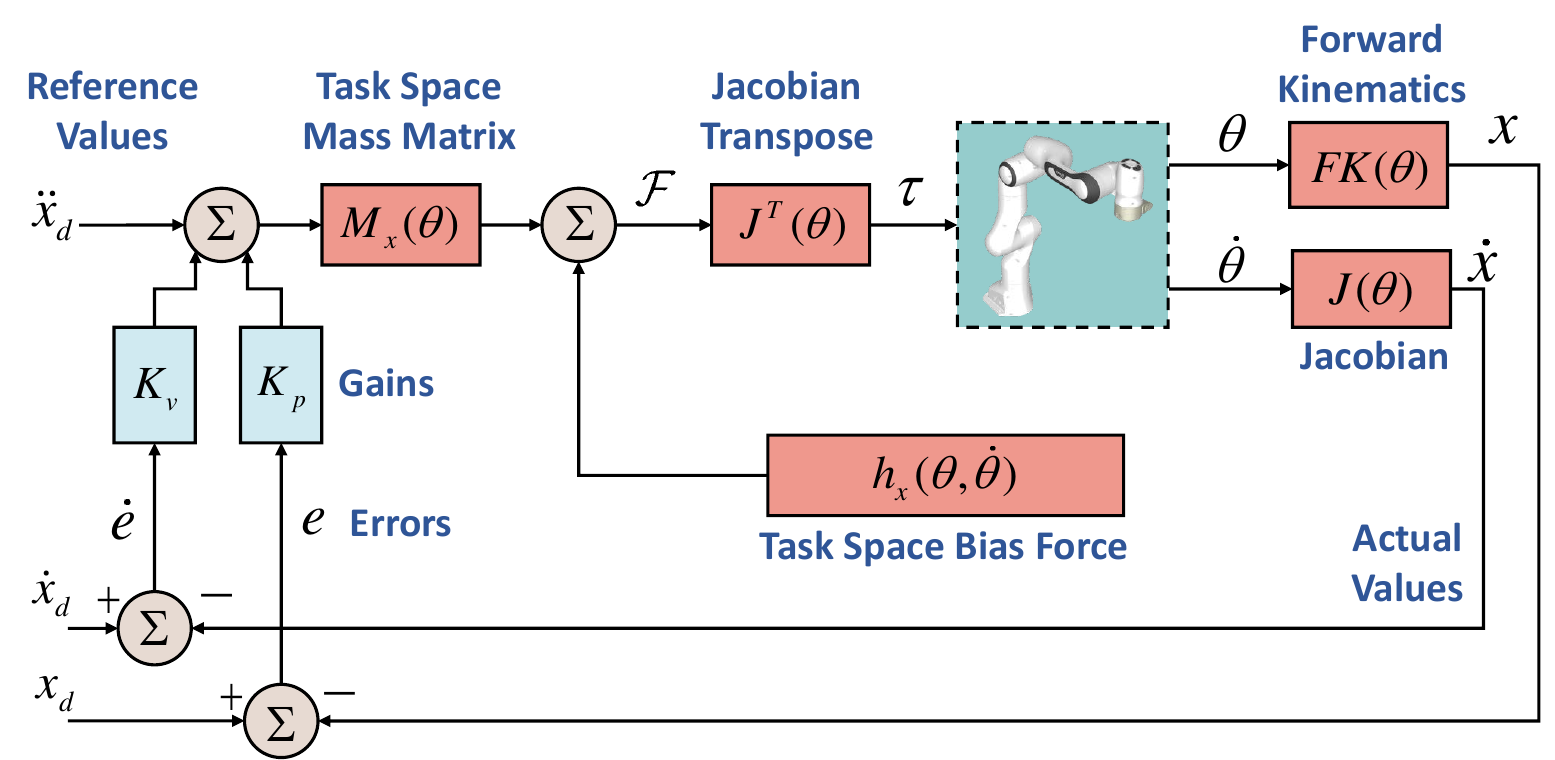}
    \caption{Task space computed torque control. }
    \label{fig:tsctc}
\end{figure}

\paragraph{Workflow.} The task space computed torque control (TS-CTC) method is widely used in robotics for precise manipulation tasks due to its ability to handle reference inputs in the task space~\cite{murray2017mathematical}. We show the control framework in Fig.~\ref{fig:tsctc}. 

\begin{equation}
\begin{aligned}
    \tau = J^T(\theta)[M_x&(\theta)(\Ddot{x}_d+K_pe+K_v\dot{e})+h_x(\theta, \dot{\theta})]\\
    e &= x_d - x \quad \dot{e} = \dot{x}_d - \dot{x}
    % \tau = J^T(\theta)[M_x(\theta)(\Ddot{x}_d+K_v(\dot{x}_d-\dot{x})+K_p(x_d-x))+h_x(\theta, \dot{\theta})]
    \label{equ:tsctc}
    \end{aligned}
\end{equation}

The input of TS-CTC has two parts. The first part is the reference trajectory $x_d$, the first order derivative $\dot{x}_d$ (velocity) and the second order derivative $\Ddot{x}_d$ (acceleration) of the reference trajectory. The second part is the joint angles $\theta$ and joint angular velocities $\dot{\theta}$ of the robot arm from the sensors. The output is the joint torque $\tau$. We describe the control process in Equ.~\ref{equ:tsctc}. Real-time robotic control, especially for smooth manipulation, necessitates high-frequency updates to joint torques. Generating and applying these torques at a frequency of 100 Hz is crucial for responsiveness and avoids jerky motion \cite{yang2023dadu, dantec2021whole}. %To achieve smooth robot control, the frequency of generating torques should be at least 100 Hz~\cite{yang2023dadu, dantec2021whole}.

%The input to TS-CTC comes from two parts. The first part is the reference trajectory $x_d$, the first order derivative $\dot{x}_d$ (velocity) and the second order derivative $\Ddot{x}_d$ (acceleration) of the reference trajectory from the upper level planning pipeline. The second part is the joint angles $\theta$ and joint angular velocities $\dot{\theta}$ of the robotic manipulator from the sensors. The output of TS-CTC is the torques applied to the joints of the manipulator. Equ.~\ref{equ:tsctc} formulates the TS-CTC. 

\paragraph{Key Computing Blocks.} TS-CTC contains five key computing blocks, which are the most computationally intensive part of the whole process. We show them as red blocks in Fig.~\ref{fig:tsctc}. The forward kinematics block calculates the pose $x$ of the end-effector in the task space based on the joint angles $\theta$. The Jacobian block calculates the Jacobian matrix $J(\theta)$ and the velocity $\dot{x}$ of the end-effector in the task space based on the joint angles $\theta$ and velocities $\dot{\theta}$. The task space mass matrix block computes the inertial matrix $M_x(\theta)$ of the robot arm in the task space based on the joint angles $\theta$. The task space bias force block computes the bias force $h_x(\theta, \dot{\theta})$ applied to the robot arm in the task space based on the joint angles $\theta$ and velocities $\dot{\theta}$.
%In the TS-CTC scheme, the forward kinematics block $FK(\theta)$ and Jacobian block $J(\theta)$ initially determine the trajectory $x$ and velocities $\dot{x}$ of the end-effector in the task space. The differences $e$, $\dot{e}$ between these calculated values $x$, $\dot{x}$ and the respective reference values $x_d$, $\dot{x}_d$ are then calculated. Amplified by their respective gains $K_p$, $K_v$, these errors $e$, $\dot{e}$ augment the desired acceleration command $\Ddot{x}_d$. After that, utilizing the task space mass matrix $M_x(\theta)$, the motion related control signals are mapped to the force domain. In combination with the external forces $\mathcal{F}^{ext}$ and the Coriolis effect $h_x(\theta, \dot{\theta})$, the aggregated force demands $\mathcal{F}$ are converted back into the joint space by means of a transposition of the Jacobian matrix $J^T(\theta)$, which generates the joint torques $\tau$, the output of the TS-CTC.\paragraph{Task space computed torque control.} The task space computed torque control (TS-CTC) method is widely used in robotics for precise manipulation tasks due to its ability to handle reference inputs in the task space. Equ.~\ref{equ:tsctc} formulates the TS-CTC. 

\subsection{\textsc{Corki}\xspace Hardware}
\label{sec:sec:break}

%\begin{table}[t]
%\centering
%\caption{The operators contained in the key computing blocks.}
%\resizebox{\columnwidth}{!}{
%\renewcommand*{\arraystretch}{1}
%\renewcommand*{\tabcolsep}{2pt}
%\begin{tabular}{|c|c|c|c|c|}
% \toprule[0.15em]
%\hline
%\diagbox{Blocks}{Operators} & \makecell[c]{Pose of 
% \\ Each Link \\ $T_i$} & \makecell[c]{Velocity of 
% \\ Each Link \\ $V_i$} & \makecell[c]{Acceleration of 
% \\ Each Link \\ $A_i$} & \makecell[c]{Force of 
% \\ Each Link \\ $F_i$} \\
% \midrule[0.1em]
%\hline
%\makecell[c]{Forward\\ Kinematics} & $\checkmark$ &  &  & \\\hline
%Jacobian & $\checkmark$ &  & & \\\hline

%\makecell[c]{Task Space \\ Mass Matrix}  & $\checkmark$ &  &  & \\\hline
%\makecell[c]{Task Space \\ Bias Force } & $\checkmark$ & $\checkmark$ & $\checkmark$ & $\checkmark$\\\hline

% \bottomrule[0.15em]
%\end{tabular}
%}
%\label{tbl:operator}
%\end{table}
\begin{figure}[t]
    \centering
    \includegraphics[width=\columnwidth]{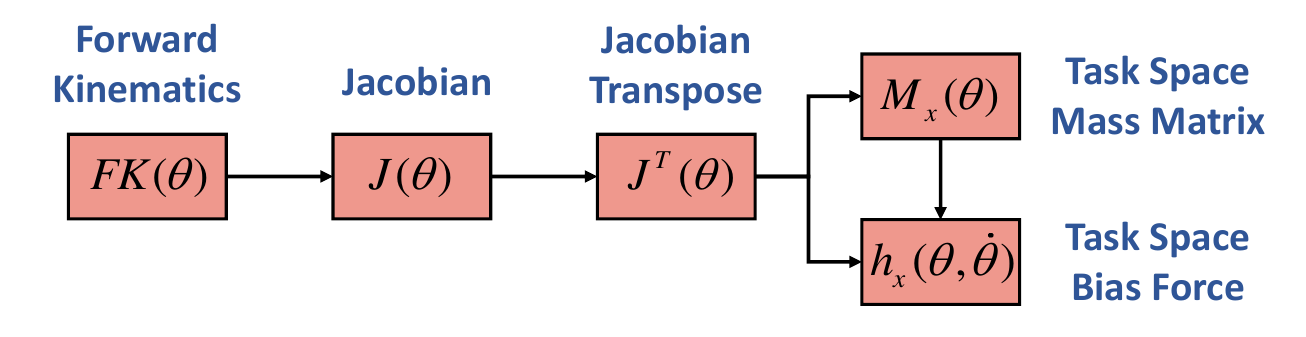}
    \caption{The results of the key blocks are reusable, as indicated by the arrows. Later stages consume partial results from early stages.}
    \label{fig:dependence}
\end{figure}

\paragraph{Bottleneck Characterization.} We analyze the compute patterns of the above control algorithms and identify two key characteristics. First, as shown in Fig.~\ref{fig:dependence}, a significant amount of intermediate data is reusable. For instance, the calculation of the Jacobian matrix reuses results from forward kinematics. Similarly, the computation of the mass matrix and bias force reuses results from the Jacobian matrix and its transpose. Second, all blocks primarily consist of four basic operations: computing the pose of each link, the velocity of each link, the acceleration of each link, and the force of each link. Due to physical laws (e.g., acceleration is the derivative of velocity), these operations follow fixed data dependencies. For example, the velocity operator consumes a six-dimensional vector from the pose operator to calculate a six-dimensional vector representing velocity. A similar trend exists between the acceleration and force operators.

%During the development of robotic manipulator applications, programmers often call library functions (the red blocks in Fig.~\ref{fig:tsctc}), e.g., forward kinematics function and Jacobian function~\cite{}. While these function calls considerably ease the programming process, their inherent computational redundancies limit the efficiency of software execution. Fig.~\ref{fig:computation} illustrates the duplicate computations among the blocks (functions). For instance, all four blocks in Fig.~\ref{fig:computation} need to calculate the pose of each link. Moreover, the task space mass matrix block and the task space bias force block require the results from the Jacobian block. 

\begin{figure}[t]
    \centering
    \includegraphics[width=\columnwidth]{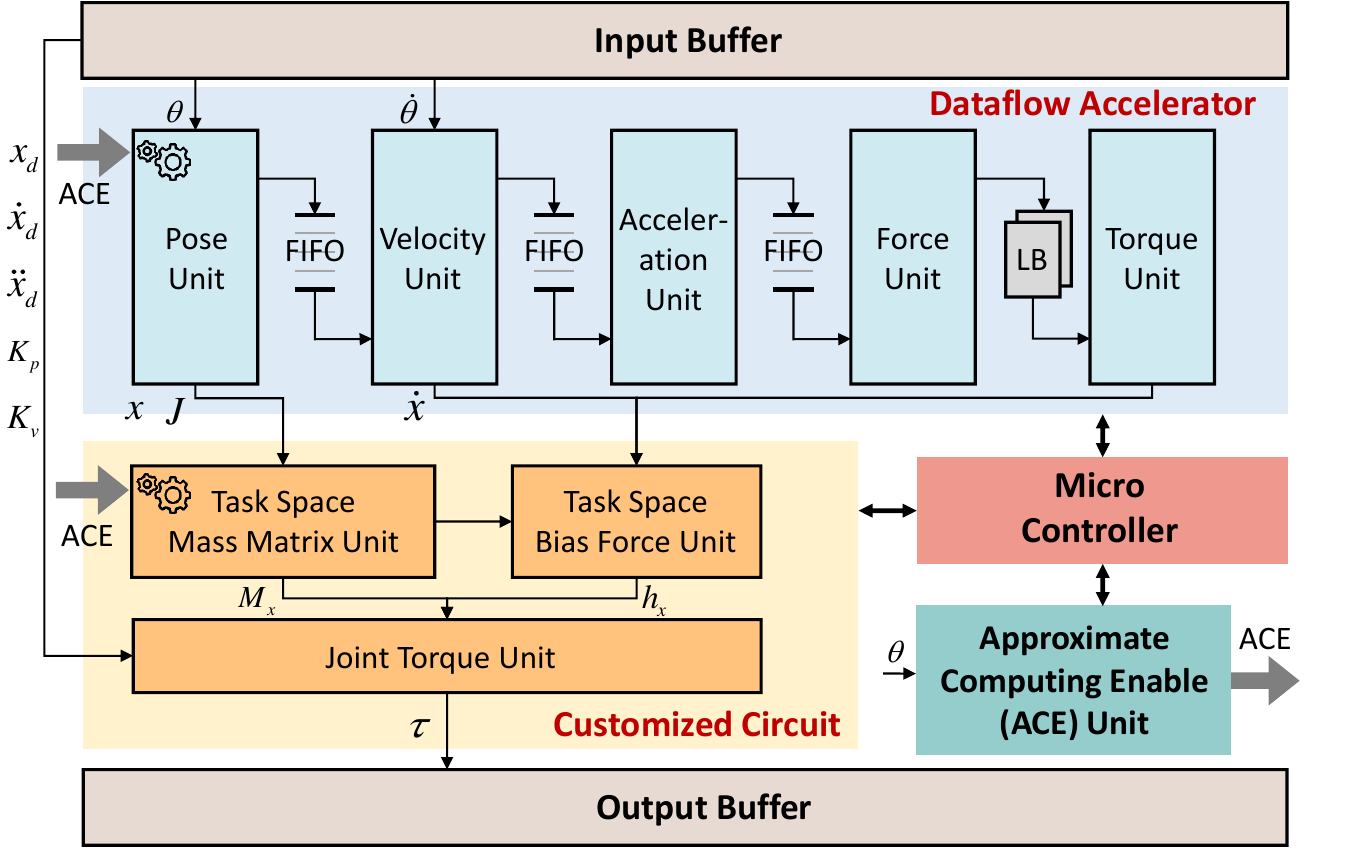}
    \caption{Hardware architecture for efficiently solving TS-CTC. Blocks in blue are in the format of a dataflow accelerator, and blocks in yellow are customized circuits. The gear indicates that the block retains the use of previously obtained results when in approximate computation mode.}
    \label{fig:hw_arch}
\end{figure}

\paragraph{Hardware Architecture.} Leveraging the above analysis, our hardware design has two goals. First, we aim to customize circuits and data pipelines to maximize intermediate data reuse, achieving high parallelization and performance. Second, we focus on customizing on-chip SRAM design to enable single read and write operations during computation, eliminating extra memory accesses.

Fig.~\ref{fig:hw_arch} shows the \textsc{Corki}\xspace architecture, which consists of two parts. The blue blocks form a dataflow accelerator, where all main operators are connected through three FIFOs and a line buffer (LB). The yellow blocks are customized circuits. A simple micro-controller manages the control flow of the accelerator.

\paragraph{Data Reuse Strategy.} According to Fig.~\ref{fig:dependence}, if each module independently computed its results, there would be significant redundant computation. For instance, the forward kinematics block calculates the pose of each link, which is also needed in the Jacobian and mass matrix computations. To mitigate this, we eliminate redundant calculations by centralizing shared computations, such as link poses, velocities, accelerations, and forces, as illustrated in the data flow accelerator of Fig.~\ref{fig:hw_arch}. Instead of re-computing values for each module, \textsc{Corki}\xspace hardware computes these quantities once and shares them across relevant computations. For example, the task space mass matrix unit utilizes precomputed poses and Jacobian matrices, while the task space bias force unit reuses precomputed torque and mass matrices, as shown in the customized circuit of Fig.~\ref{fig:hw_arch}.  Since the Jacobian matrix and its transpose are required multiple times during a single control computation cycle, and given that the Jacobian matrix is  small (typically at most $6\times7$ in robotic manipulation tasks~\cite{murray2017mathematical}), we allocate a separate memory copy for the Jacobian transpose. This avoids irregular memory access patterns that could otherwise lead to access conflicts, ensuring more efficient processing. Experimental results demonstrate that using our data reuse strategy results in a $54.0\%$ reduction in total latency.

\paragraph{Pipeline Design.} In the data flow accelerator, the computation of pose, velocity, acceleration, and force must follow a sequential order due to physical constraints, e.g., velocity is the derivative of position and acceleration is the derivative of velocity. Thus, computing the acceleration of a link requires knowing its pose and velocity. However, this dependency exists only within the same link. We find that computations for different links can proceed in parallel. For example, while computing link 1’s force, we can simultaneously compute link 2’s acceleration and link 3’s velocity. This pipelined design significantly reduces the delay for computing forces and torques across all links. Experimental results show that our pipelined design results in a further $69.6\%$ reduction in total latency based on the use of the data reuse strategy. When compared to the baseline hardware implementation without any optimizations (i.e., lacking both the data reuse strategy and pipeline design), the total latency is reduced by $86.0\%$.
% Fig.~\ref{fig:hw_arch} shows the \textsc{Corki}\xspace architecture, which consists of two parts. The blue blocks form a dataflow accelerator, where all main operators are connected through three FIFOs and a line buffer (LB). This design enables extreme pipelining; for example, the velocity calculation of the first link can start while the pose calculation of the second link begins. The yellow blocks are customized circuits, with the task space mass matrix unit reusing data from the pose unit and the task space bias force unit reusing data from both the velocity unit and the torque unit. There are occasional stalls in the accelerators due to differing latencies between the dataflow accelerator and the customized circuits. A simple micro-controller manages the control flow of the accelerator.

\paragraph{Memory Optimization.} Our on-chip buffer design is highly effective. In the first four stages of the dataflow accelerator, three FIFOs store intermediate data, as the producer and consumer rates are identical. A line buffer between the force unit and the torque unit captures the rate mismatch between them. The remaining intermediate data is stored in a small scratchpad memory. This combination of different on-chip buffer designs allows for minimal on-chip SRAM consumption while ensuring no data communication with off-chip DRAM during execution.

\subsection{Application-specific Approximate Computing}
\label{sec:sec:dyna}

\paragraph{Opportunity.} We observe that robotic control has a unique feature: the compute frequency is high, yet the change in each control signal is low. For a 7-DoF robot arm, the movement in each joint is minimal each time. However, the computation of control signals is based on joints, as illustrated in the previous section. A joint-based approximation is possible to further save computation and reduce latency.

\paragraph{Quantitative Analysis.} To quantitatively demonstrate our observation, we perform an experiment. We use a 7-DoF Franka Emika Panda robot arm~\cite{gaz2019dynamic} and monitor the item-wise changes in the mass matrix while slightly adjusting each joint by an angle. For example, we first record all the items in the mass matrix, then change the first joint by 0.1 radians (approximately 6 degrees), 0.3 radians (approximately 17 degrees), and 0.5 radians (approximately 29 degrees), monitoring the changes in the mass matrix. We repeat the same experiments for all the joints in the robot arm.

\begin{figure}[t]
    \centering
    \includegraphics[width=1\columnwidth]{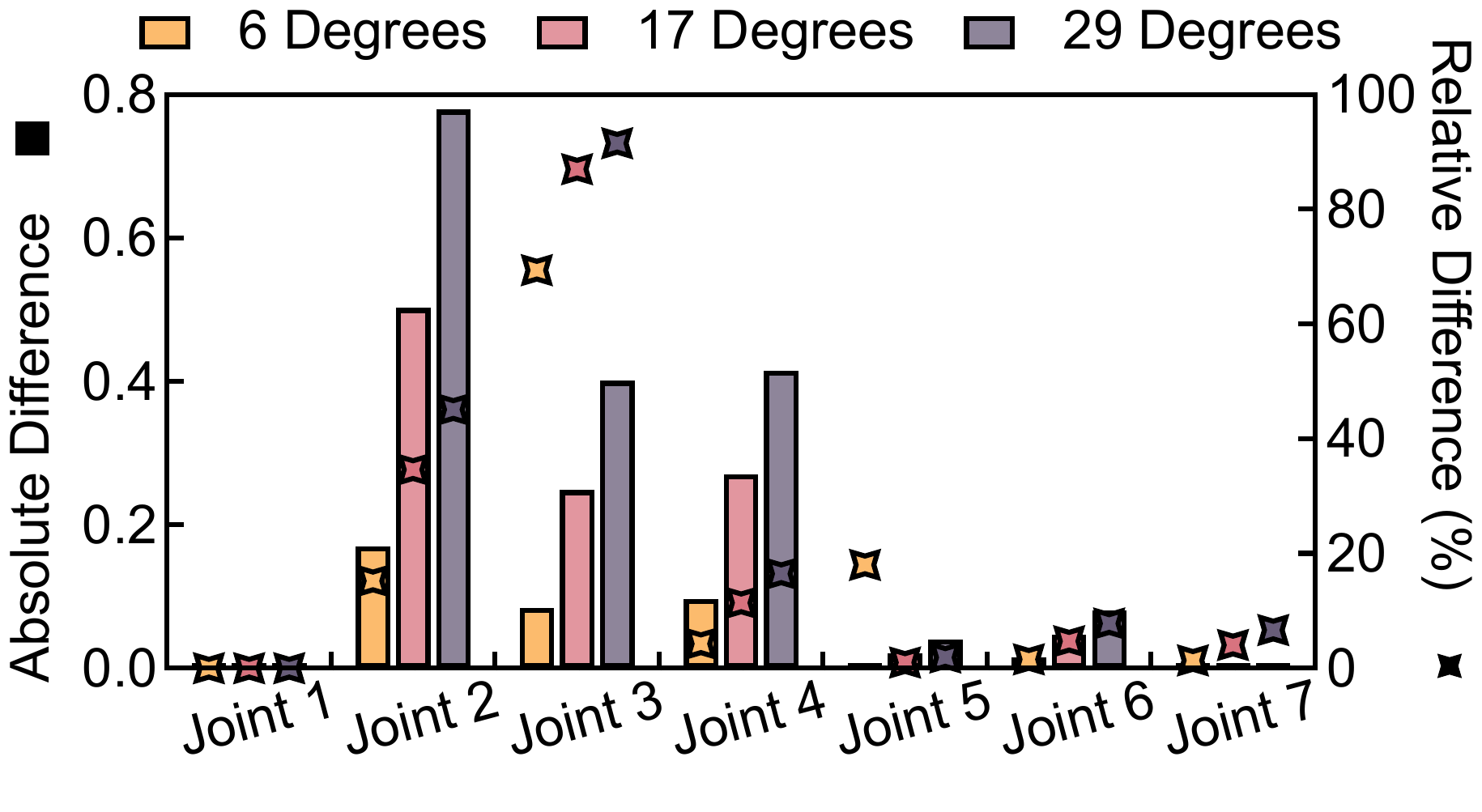}
    \caption{The maximum difference in the elements of the mass matrix before and after movements in the joint. The experiments are conducted on a Franka Emika Panda robot arm. The movement consists of rotation with angles of 6 degrees, 17 degrees and 29 degrees on all 7 joints. }
    \label{fig:mass_matrix_variation}
    \vspace{-8pt}
\end{figure}

\begin{figure}[t]
    \centering
    \includegraphics[width=0.85\columnwidth]{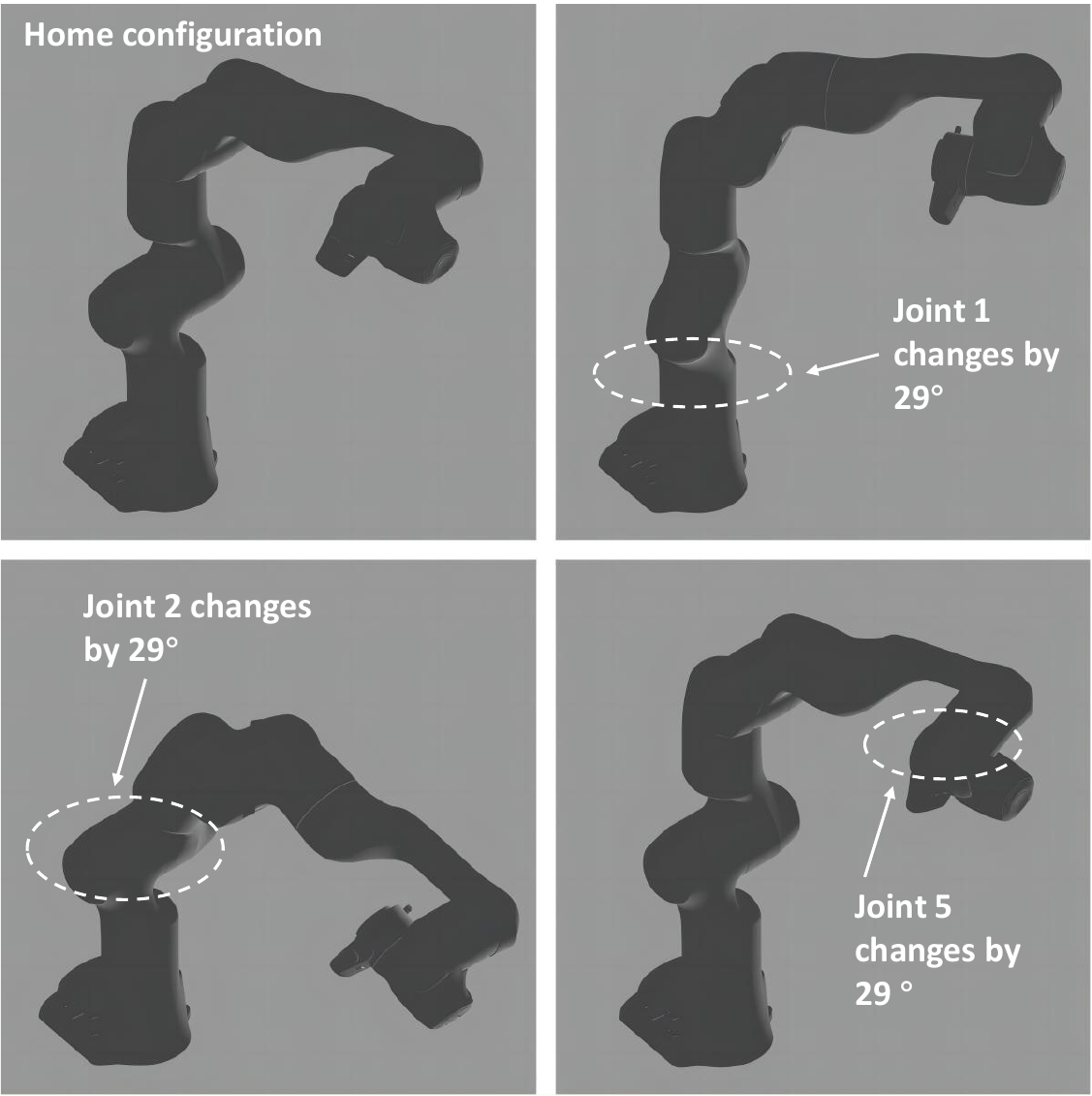}
    \caption{The morphology of the Franka Emika Panda robot arm in different configurations. We change joint 1, joint 2 and joint 5 by 29 degrees and show the difference.}
    \label{fig:panda}
\end{figure}

We show the results in Fig.~\ref{fig:mass_matrix_variation}. The results indicate that when motion occurs in joints 1 and 7, the mass matrix remains nearly constant. This phenomenon is illustrated in the top right and bottom right figures in Fig.~\ref{fig:panda}. Movements in the end joints (joint 1 and joint 7) have minimal impact on the morphology of the robot arm, leading to less significant changes in the mass matrix. Similarly, for joints 5 and 6, the maximum variation in matrix elements does not exceed 0.1 even with an angular change of 29 degrees.

However, the situation is different for the joints in the middle of the robot arm. When joint 2 moves, even a change of 6 degrees results in a maximum absolute change in matrix elements of 0.17 (with a maximum relative change of approximately 15.4\%). When the motion increases to 29 degrees, the maximum relative change in elements can be as high as 45.2\%. The bottom left figure in Fig.~\ref{fig:panda} shows that when the middle joints undergo movement, the morphology of the robot arm is significantly changed, necessitating the re-computation of all parameters in the control process.

%Similarly in joint 5 and joint 6, the item in mass matrices only change by less than \fixme{20\%} when the change is 0.5 radians. 

%\paragraph{Analysis.} To illustrate this, we analyze the changes in the mass matrices of the Franka Panda manipulator~\cite{} which we use when the joint angles undergo slight variations. Fig.~\ref{fig:mass_matrix} illustrates the maximum variations in the elements of the mass matrices before and after adjustments in the joint angles of the manipulator. For instance, the left bar of the 4-th joint in Fig.~\ref{fig:mass_matrix} shows that when the 4-th joint angle of the manipulator changes by 0.1 radians (approximately 6 degrees), the maximum variation in the elements of the mass matrices is only about 0.1. As evident from Fig.~\ref{fig:mass_matrix}, changes in the joint angles located in the middle of the manipulator (the 2-nd to 4-th joints) have a more significant impact on the mass matrix compared to those at the ends of the manipulator (the 1-st joint and the 5-th to 7-th joints).

%Fig.~\ref{fig:panda} intuitively explains the reason for the above phenomenon. The top-left portion of Fig.~\ref{fig:panda} shows the Franka Panda robotic manipulator in its home configuration. The top-right portion illustrates the manipulator with its 1-st joint rotated by 0.5 radians (approximately 29 degrees). The bottom-left portion displays the manipulator with its 2-nd joint rotated by 0.5 radians, and the bottom-right portion shows the manipulator with its 5-th joint rotated by 0.5 radians. Changes in the joint angles located in the middle of the manipulator (e.g., the 2-nd joint) significantly affect its morphology, leading to greater variations in the mass matrix. In contrast, changes in the joint angles at the ends of the manipulator (e.g., the 5-th joint) have minimal impact on its morphology, resulting in only slight variations in the mass matrix.

\paragraph{Approximate Computation.} We design a simple yet effective approximate computing method to dynamically update the control parameters, reducing the computational costs in the control process. Whether or not to perform approximate computations is evaluated by the hardware, as shown in Fig.~\ref{fig:hw_arch}. Specifically, given the input $\theta$, we first compute the probability of each matrix (e.g., Jacobian matrix, mass matrix, etc.) needing an update based on an impact factor derived from the angular movement of each joint. In this process, the joints with a small impact on parameter changes have smaller impact factors, while the joints with a large impact on parameter changes have larger impact factors. The probability computation consumes less than 100 FLOPs, which does not affect the final latency.

If the probability of updating a matrix exceeds a certain threshold, the corresponding computation to generate that matrix is performed. Otherwise, the corresponding elements from the previous control cycle are reused. We observe that over 51\% of matrix updates can be avoided without any loss in control accuracy, using trajectory error as the metric.

%Given the above observations, we do not adopt a simple approach that only update the joint angle-based computational blocks when the joint angles exceeded a single threshold. Instead, we assigned specific weights to each joint angle. Joints that significantly impacted the model parameters (e.g., mass matrix) are given a higher weight, consequently updating these blocks when these joint angles exceeded a smaller threshold. For joints that have a minimal impact on the model parameters, updates are only triggered when their angles exceeded a larger threshold.

\subsection{System Pipeline}
\label{sec:arch:pipe}
There are three key components in the system we propose. First, network inference that happens on the server will predict the trajectory. The parameters of the trajectory will be sent to the controller, which is located on the robot. The controller calculates the high-frequency actual control signals to enable the robot to move as the trajectory plans, and the robot will move according to the control signals. During the movement of the robot, at random time steps before the trajectory ends, images will be captured by the camera mounted on the robot. These images will be sent back to the server while the robot continues to finish the rest of the trajectory. Thus, communication and robotic control can be executed in parallel. When the robot reaches the end of the trajectory, it will capture another image and send it back to the server. A new trajectory will be predicted through the LLM inference using this image and previous images. 

\section{Experimental Methodology}
\label{sec:setup}

\begin{sloppypar}
This section describes our evaluation methodology. First, we will discuss the experimental setup, including the software, dataset, and hardware (Sec.~\ref{sec:setup:setup}). Then, we will cover the baselines we compare and the variations of \textsc{Corki}\xspace (Sec.~\ref{sec:setup:base}). 
\end{sloppypar}

\subsection{Experimental Setup}
\label{sec:setup:setup}

We build \textsc{Corki}\xspace on the foundation of RoboFlamingo~\cite{livision}, but our work is extensible to other action-prediction-based embodied AI robots. We implement the algorithm innovation in PyTorch~\cite{paszke2019pytorch}, where the network output predicts a trajectory instead of a discrete action. This predicted trajectory is then fed back into the simulation environments to test the robot's task completion capabilities.

We use the Calvin~\cite{mees2022calvin} dataset and software simulation environments, one of the most widely used embodied AI datasets. Calvin includes 34 different tasks with 22994 demonstrations for training and 1000 sequences for testing. We evaluate our algorithm in two different scenarios: Seen scenarios, where the tasks in the testing set are similar but not identical to those encountered during training, and unseen scenarios, which are more challenging as the tasks are entirely new and have not been encountered during training.

\paragraph{Tasks and Metrics.} The tasks are categorized into five types: moving an object, turning a switch on and off, pushing and pulling a drawer, rotating an object, and lifting an object. We use two metrics to evaluate the algorithm's accuracy: success rate and average job length. The success rate is the most straightforward metric for quantifying a single task, calculated as the number of successful sequences divided by the total sequences. Given that the embodied AI algorithms are designed to improve robots' abilities on long-horizon jobs, we further report the accuracy of finishing a job. Each job contains five consecutive tasks. The average job length measures how many tasks the robot can complete within a job, with a maximum of 5.

\paragraph{Trajectory Comparison.} We further utilize two different metrics to illustrate why the results we predict are better:
\begin{itemize}
    \item Mean trajectory error. We compare the geographic distance between the predicted trajectory and the ground truth, using root mean square error (RMSE) as the metric. Generally, a smaller RMSE indicates better robot trajectory.
    \item Maximum trajectory distance. We also compare the maximum distance between the predicted and ground truth trajectories. A larger maximum distance denotes a higher likelihood of failure. 
\end{itemize}

\paragraph{Hardware.} Inference latency and energy consumption are measured on an Nvidia V100 GPU, with power readings obtained via NVML~\cite{vonnvml}. We measure latency and energy consumption for control algorithms running on an Intel Core i7-6770HQ CPU. We implement \textsc{Corki}\xspace hardware on a Xilinx Zynq-7000 SoC ZC706 FPGA~\cite{zc706} to assess real hardware performance. Additionally, we establish Wi-Fi communication between a 7-DoF Franka Emika Panda robot arm~\cite{gaz2019dynamic} and our server to measure communication latency. 

The entire evaluation is conducted in this manner. For the baseline, each frame undergoes three stages, LLM inference (real-world latency measured on the server), communication (real-world latency measured through sending actual frames from the robot to the server), control (real-world latency measured on the processor of the robot, which is the Intel Core i7-6770HQ CPU). For \textsc{Corki}\xspace, each trajectory undergoes three stages, LLM inference generating trajectory (real-world latency measured on the server), communication (real-world latency measured through sending actual frames from the FPGA to the server), control (real data measured on our FPGA board with real trajectories as inputs). We use synchronized timestamps to accurately measure the communication latency.

\subsection{Baselines and Variations}
\label{sec:setup:base}
\paragraph{Baselines.} We train RoboFlamingo using the Calvin dataset for accuracy comparison. The results are either higher or equivalent to the reported version. For latency and energy consumption comparisons, we establish a baseline using the traditional execution pipeline of embodied AI algorithms, where the inference latency, control latency, and communication latency are accumulated in each frame.

\paragraph{Variations.} As discussed earlier, \textsc{Corki}\xspace can predict the trajectory of the next $N$ steps, with each step taking approximately 3.3 ms. Given the predicted trajectory covering $N$ steps, the robots can take anywhere from 1 step to up to $N$ steps. Longer steps reduce the inference frequency but may also lead to lower accuracy. In our evaluation, we predict nine steps each time and vary the steps taken from 1 to 9 with a stride of 2, creating five variations named \underline{\textsc{Corki}\xspace-T}, where T represents the actual steps taken.

In addition to the fixed step variations, we evaluate adaptive options as discussed in Section \ref{sec:algo:adap}. We name this variation \underline{\textsc{Corki}\xspace-ADAP}. In \underline{\textsc{Corki}\xspace-ADAP}, the robot's steps are selected by the waypoints identification algorithm and are smaller than $N$.

To demonstrate the effectiveness of the accelerator, we introduce a variant named \underline{\textsc{Corki}\xspace-SW}, which employs trajectory prediction with five steps (as in \underline{\textsc{Corki}\xspace-5}) while retaining the baseline CPU for control processing. It applies same approximation with \textsc{Corki}\xspace-5. %named \underline{\textsc{Corki}\xspace-SW}, in this case, only the software trajectory prediction is applied with the steps equal to 5 (the same as in \underline{\textsc{Corki}\xspace-5}), yet the control process still runs on the baseline CPU. 

\begin{table}[t]
\centering
\caption{Accuracy on seen tasks. Baseline is retrained.}
\resizebox{1\columnwidth}{!}{
\begin{tabular}{cccccccccc}
\hline
\multirow{2}{*}{Variation} & \multicolumn{5}{c}{Task Completed in a Sequence} & \\
& 1 & 2 & 3 & 4 & 5 & Avg Len\\
\hline
\rowcolor{mygray}
RoboFlamingo& 89.5\% & 71.9\% & 55.6\% & 43.4\% &  31.2\%& 2.916\\
\textsc{Corki}\xspace-1& 89.1\% & 75.3\% & 59.2\% & 47.1\% & 37.1\% & 3.078 \\
\textsc{Corki}\xspace-3& 89.4\% & 75.7\% & 62.6\% & 52.9\% & 42.8\% & 3.234 \\
\textsc{Corki}\xspace-5& 92.3\% & \textbf{\underline{80.0\%}} & \textbf{\underline{67.4\%}} & \textbf{\underline{56.6\%}} & \textbf{\underline{45.8\%}} & \textbf{\underline{3.421}}\\
\textsc{Corki}\xspace-7&89.1\% & 73.8\% & 59.5\% & 48.7\% & 38.1\% & 3.092 \\
\textsc{Corki}\xspace-9& 88.0\% & 72.0\% & 56.4\% & 46.3\% & 35.6\% & 2.983 \\
\textsc{Corki}\xspace-ADAP & \textbf{\underline{93.5\%}} & 77.7\% &  61.4\% & 49.1\% & 38.3\% & 3.2 \\
\textsc{Corki}\xspace-SW (using \textsc{Corki}\xspace-5)& 92.3\% & \textbf{\underline{80.0\%}} & \textbf{\underline{67.4\%}} & \textbf{\underline{56.6\%}} & \textbf{\underline{45.8\%}} & \textbf{\underline{3.421}} \\
\hline
\label{D-D}
\end{tabular}}
\end{table}

\begin{table}[t]
\centering
\caption{Accuracy on unseen tasks. Baseline is retrained.}
\resizebox{1\columnwidth}{!}{
\begin{tabular}{cccccccccc}
\hline

\multirow{2}{*}{Variation} & \multicolumn{5}{c}{Task Completed in a Sequence} & \\
& 1 & 2 & 3 & 4 & 5 & Avg Len\\

\hline
\rowcolor{mygray}
RoboFlamingo& 82.4\% & 61.9\% & 46.6\% & 33.1\% & 23.5\% & 2.48 \\
\textsc{Corki}\xspace-1& 86.0\% & 68.0\% & 52.6\% & 40.3\% & 30.0\% & 2.769 \\
\textsc{Corki}\xspace-3& 83.2\% & 65.6\% & 50.7\% & 37.2\% & 27.5\% & 2.642 \\
\textsc{Corki}\xspace-5 & \textbf{\underline{85.9\%}} & 68.4\% & \textbf{\underline{54.3\%}} & \textbf{\underline{42.2\%}} & \textbf{\underline{31.6\%}} & 2.824\\
\textsc{Corki}\xspace-7&83.8\% & 65.5\% & 50.5\% & 40.6\% & 31.9\% & 2.723 \\
\textsc{Corki}\xspace-9& 79.4\% & 59.5\% & 44.0\% & 33.7\% & 24.7\% & 2.413\\
\textsc{Corki}\xspace-ADAP & 85.7\% & \textbf{\underline{69.4\%}} & 54.1\% & 41.9\%  & 31.6\% & \textbf{\underline{2.827}} \\
\textsc{Corki}\xspace-SW (using \textsc{Corki}\xspace-5)& \textbf{\underline{85.9\%}} & 68.4\% & \textbf{\underline{54.3\%}} & \textbf{\underline{42.2\%}} & \textbf{\underline{31.6\%}} & 2.824\\
\hline
\label{ABC-D}
\end{tabular}}
\end{table}

\section{Evaluation}
\label{sec:eval}
We evaluate \textsc{Corki}\xspace in this section. We first show that the \textsc{Corki}\xspace  accelerator has a low hardware resource consumption (Sec.~\ref{sec:eval:area}). We then evaluate both the accuracy of \textsc{Corki}\xspace (Sec.~\ref{sec:eval:acc}) and corresponding latency and energy saving (Sec.~\ref{sec:eval:perf}).

\subsection{Hardware Resource Consumption}
\label{sec:eval:area}

%\begin{table}[t]
%\centering
%\caption{FPGA resource consumption (utilization percentages and absolute numbers).}
%\resizebox{0.5\columnwidth}{!}{
%\renewcommand*{\arraystretch}{1}
%\renewcommand*{\tabcolsep}{2pt}
%\begin{tabular}{cccc}
%\toprule
% \hline
%BRAM & DSP & FF & LUT \\ 
%\midrule
%\makecell[c]{6.6\%\\(36)} & \makecell[c]{13.6\%\\(122)} & \makecell[c]{7.8\%\\(34300)} & \makecell[c]{16.9\%\\(36929)} \\ 
%\bottomrule
%\end{tabular}
%}
%\label{tbl:resource}
%\end{table}

%The \textsc{Corki}\xspace accelerator is compact and does not require significant hardware resources, making it feasible for deployment on a real robot.
The compact \textsc{Corki}\xspace accelerator requires minimal hardware resources, making it suitable for deployment on a real robot. It consumes only 13.6\% of digital signal processors (DSP), 7.8\% of flip-flops (FF), and 16.9\% of look-up tables (LUT). The specialized on-chip buffer design is effective; the \textsc{Corki}\xspace accelerator utilizes only 6.6\% of the total block random access memory (BRAM), with no data communication with off-chip DRAM during each control process.

%The \textsc{Corki}\xspace hardware consumes 6\% of block random access memory (BRAM), 13\% of digital signal pocessors (DSP), 7\% of Flip-Flops (FF), and 16\% of Look-Up Tables (LUT), as presented in Tbl.~\ref{tbl:resource}. Our primary focus is on maximizing parallel processing and performance by employing custom circuits and efficient data pipelines, while also minimizing external memory interactions through a tailored on-chip SRAM strategy. This efficient use of resources, especially the low BRAM utilization, validates our design approach.

%we ensure high performance with minimal off-chip communication, while effectively managing latency differences through a micro-controller. The balanced consumption across various FPGA resources underscores a carefully optimized design that achieves intended functionality without excessive resource demands.

\subsection{Accuracy}
\label{sec:eval:acc}

% \begin{table}[t]
% \centering
% \caption{Accuracy on seen tasks. Baseline is retrained.}
% \resizebox{1\columnwidth}{!}{
% \begin{tabular}{cccccccccc}
% \hline
% \multirow{2}{*}{Variation} & \multicolumn{5}{c}{Task Completed in a Sequence} & \\
% & 1 & 2 & 3 & 4 & 5 & Avg Len\\
% \hline
% \rowcolor{mygray}
% RoboFlamingo& 89.5\% & 71.9\% & 55.6\% & 43.4\% &  31.2\%& 2.916\\
% \textsc{Corki}\xspace-1& 89.1\% & 75.3\% & 59.2\% & 47.1\% & 37.1\% & 3.078 \\
% \textsc{Corki}\xspace-3& 89.4\% & 75.7\% & 62.6\% & 52.9\% & 42.8\% & 3.234 \\
% \textsc{Corki}\xspace-5& 92.3\% & \textbf{80.0\%} & \textbf{67.4\%} & \textbf{56.6\%} & \textbf{45.8\%} & \textbf{3.421}\\
% \textsc{Corki}\xspace-7&89.1\% & 73.8\% & 59.5\% & 48.7\% & 38.1\% & 3.092 \\
% \textsc{Corki}\xspace-9& 88.0\% & 72.0\% & 56.4\% & 46.3\% & 35.6\% & 2.983 \\
% \textsc{Corki}\xspace-ADAP & \textbf{93.5\%} & 77.7\% &  61.4\% & 49.1\% & 38.3\% & 3.2 \\
% \textsc{Corki}\xspace-SW& 92.3\% & \textbf{80.0\%} & \textbf{67.4\%} & \textbf{56.6\%} & \textbf{45.8\%} & \textbf{3.421} \\
% \hline
% \label{D-D}
% \end{tabular}}
% \end{table}

\begin{figure}[t]
%\vspace{-10pt}
\centering
\subfloat[\small{Mean trajectory error.}]
{
  \includegraphics[trim=0 0 0 0, clip, width=0.45\columnwidth]{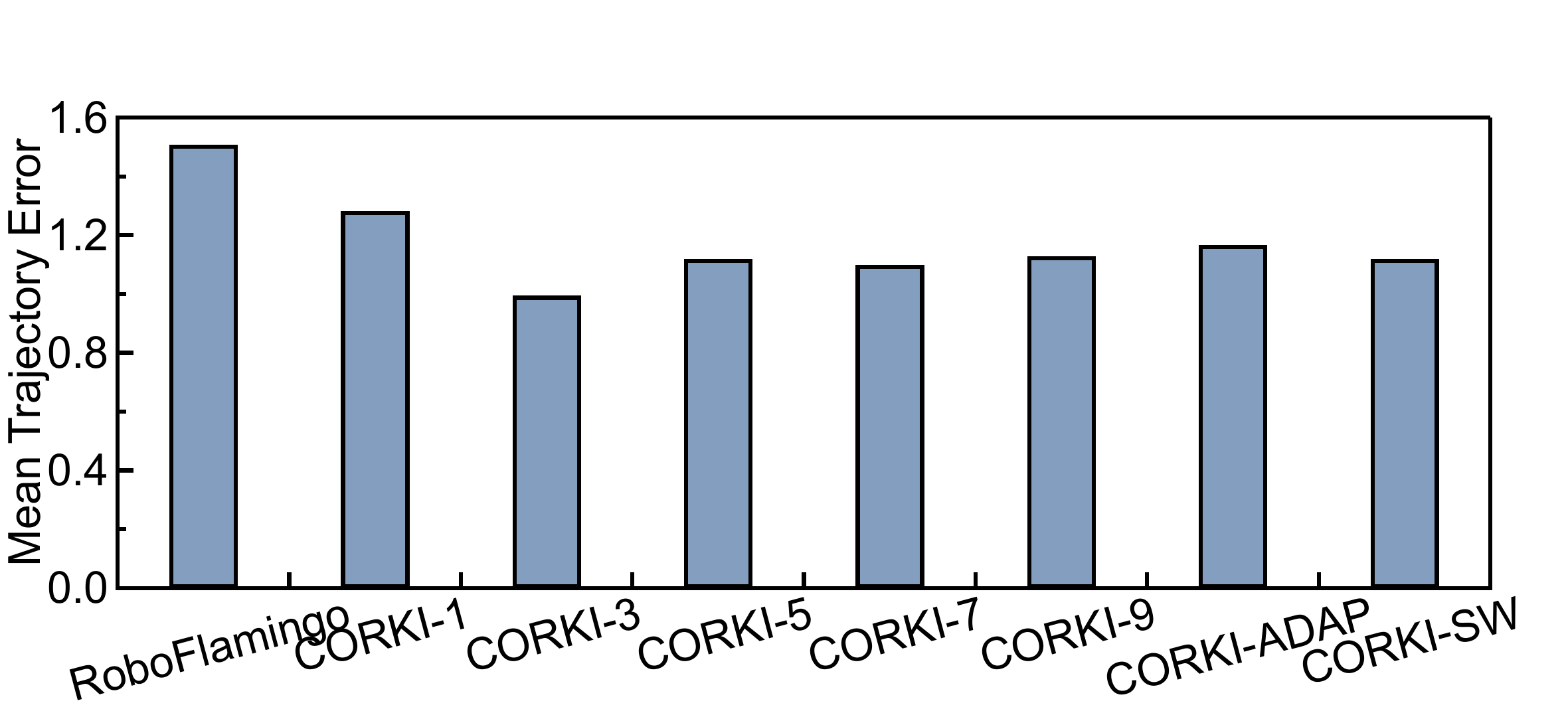}
  \label{fig:rmse}
}
%\hspace{3pt}
\subfloat[\small{Maximum trajectory distance.}]
{
  \includegraphics[trim=0 0 0 0, clip, width=0.45\columnwidth]{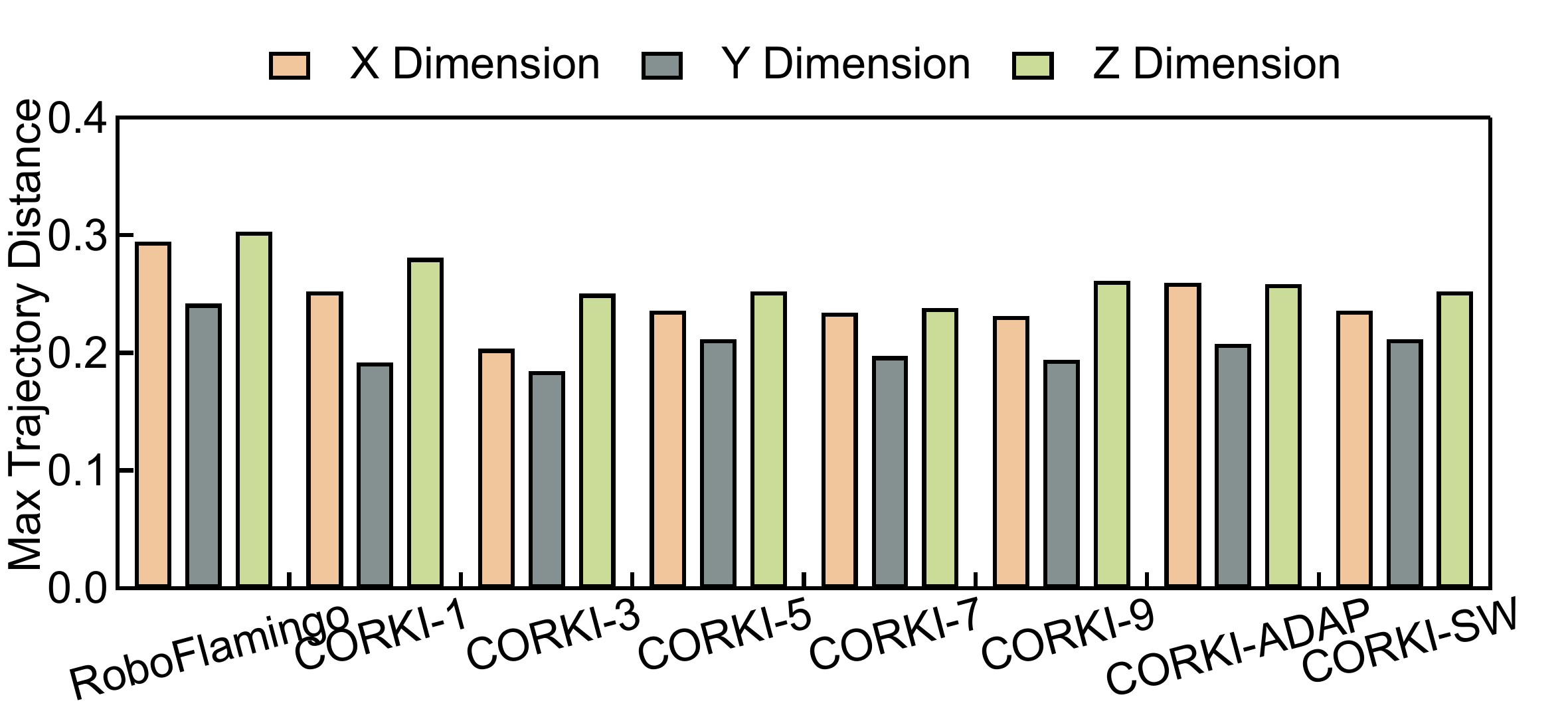}
  \label{fig:dist}
}
\caption{Trajectory comparison between \textsc{Corki}\xspace and RoboFlamingo with two quantitative metrics.}
\label{fig:trajerror}
\end{figure}

\begin{figure*}[t]
%\vspace{-10pt}
\centering
\subfloat[\small{X dimension trajectory.}]
{
  \includegraphics[trim=0 0 0 0, clip, width=0.66\columnwidth]{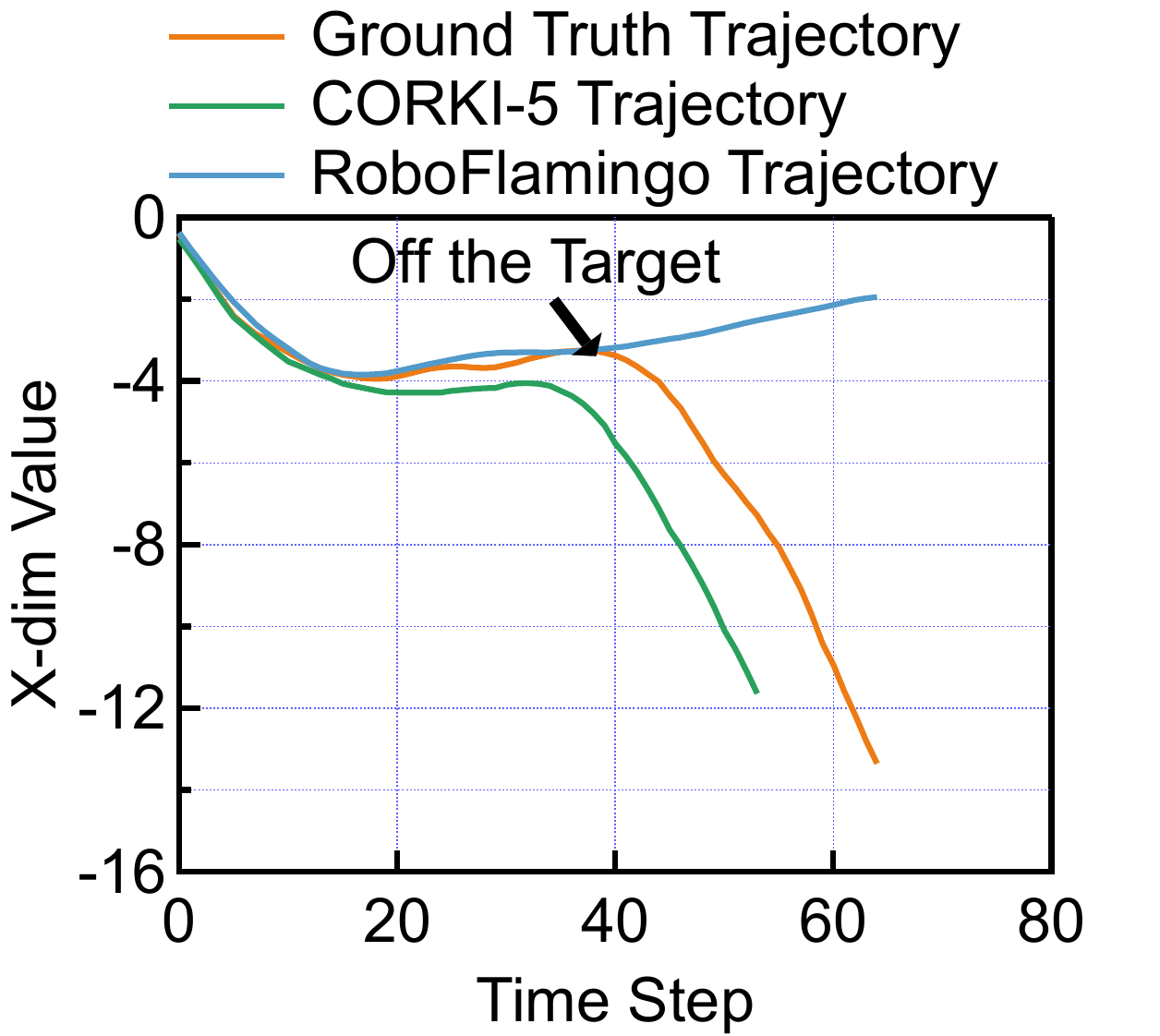}
  \label{fig:trajx}
}
%\hspace{3pt}
\subfloat[\small{Y dimension trajectory.}]
{
  \includegraphics[trim=0 0 0 0, clip, width=0.66\columnwidth]{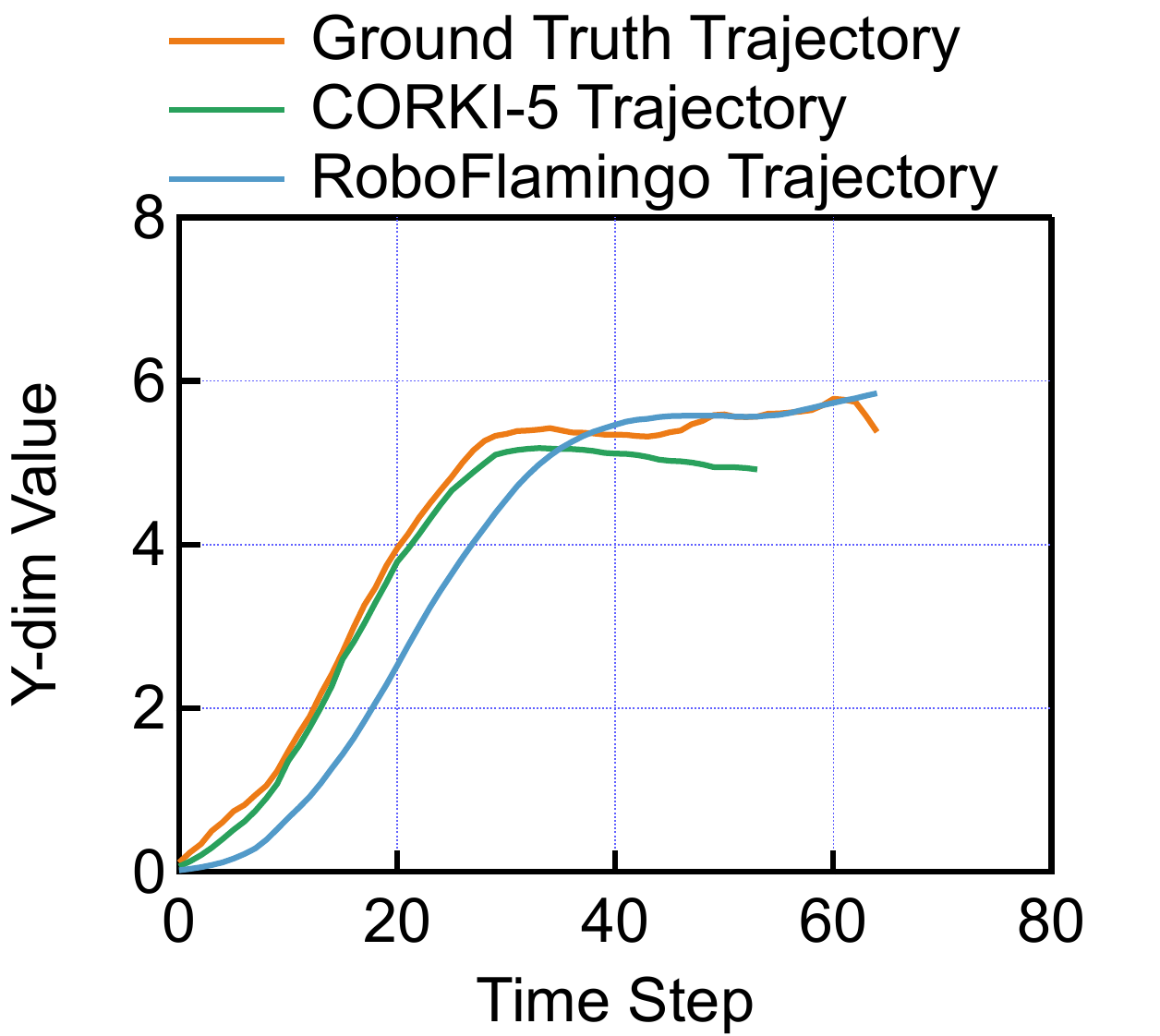}
  \label{fig:trajy}
}
\subfloat[\small{Z dimension trajectory.}]
{
  \includegraphics[trim=0 0 0 0, clip, width=0.66\columnwidth]{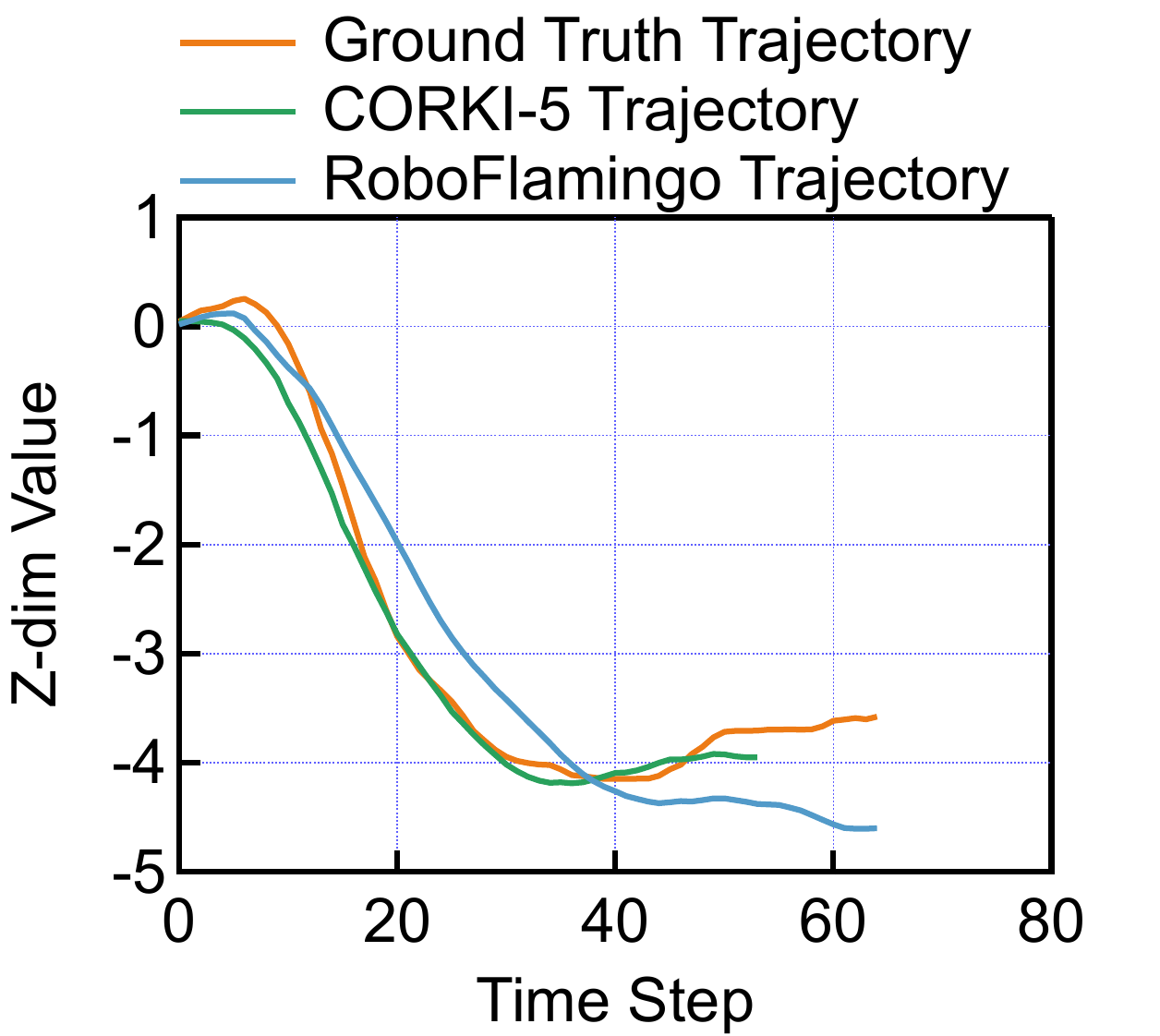}
  \label{fig:trajz}
}
\caption{Trajectory comparison of a randomly picked sequence from the test set. It is clearly shown that trajectories of \textsc{Corki}\xspace can follow the ground truth, while trajectories of Roboflamingo are off the target. We only show \underline{\textsc{Corki}\xspace-5} for simplicity.}
\label{fig:traj}
\end{figure*}

\paragraph{Success Rate and Average Job Length.} We show accuracy results on seen scenarios and unseen scenarios in Tbl.~\ref{D-D} and Tbl.~\ref{ABC-D}. Almost all variations of \textsc{Corki}\xspace outperform the baseline in terms of both success rate and average job length, except for \underline{\textsc{Corki}\xspace-9} in unseen scenarios. On average, \textsc{Corki}\xspace improves the success rate by 8.6\% and the average job length by 0.3. In unseen scenarios, these improvements are 8.1\% and 0.2, respectively.
%\textsc{Corki}\xspace improves the success rate by 8.6\% on average compared to the baseline, and improves the average job length by 0.3. These two statistics for unseen scenarios are 8.1\% and 0.2.

%We perform evaluation on seen scenarios and unseen scenarios and show the results in Tbl.~\ref{D-D} and Tbl.~\ref{ABC-D}. Almost all variations of \textsc{Corki}\xspace outperform the baseline in terms of both success rate and average job length, except for \texttt{\textsc{Corki}\xspace-9} in unseen scenarions. \textsc{Corki}\xspace improves the success rate by 8.6\% on average compared to the baseline, and improves the average job length by 0.3. These two statistics for unseen scenarios are \yiyang{8.1\%} and \yiyang{0.2}. 
\begin{sloppypar}
Among all fixed-step variations of \textsc{Corki}\xspace, \underline{\textsc{Corki}\xspace-5} achieves the highest accuracy and significantly outperforms the baseline. On seen tasks, the average job length is improved by 17.3\% compared to the baseline, with a gain of 0.5 in job length. The trend observed among all \textsc{Corki}\xspace variations is that accuracy improves as the length of the actual trajectory taken increases. However, after reaching its peak accuracy, there is a gradual degradation in performance when the length of the actual trajectory taken continues to increase.
\end{sloppypar}

\underline{\textsc{Corki}\xspace-ADAP} selects the length of the actual trajectory by identifying waypoints with significant movements. We observe that the results of \underline{\textsc{Corki}\xspace-ADAP} fall between those of \underline{\textsc{Corki}\xspace-7} and \underline{\textsc{Corki}\xspace-5} in seen tasks, and it even outperforms \underline{\textsc{Corki}\xspace-5} in unseen tasks. This demonstrates that determining length during runtime is effective. \underline{\textsc{Corki}\xspace-SW} achieves the same accuracy as \underline{\textsc{Corki}\xspace-5} because the only difference is whether the control process occurs on the accelerator, which does not affect accuracy.

\paragraph{Understanding the Results.} The improvement brought by \textsc{Corki}\xspace is significant. \textsc{Corki}\xspace outperforms the baseline in almost all cases because trajectory naturally provides a more robotic-friendly supervision during algorithm training. When the datasets of embodied AI algorithms are constructed, the collection of the ground truth was in the form of trajectory at first. In contrast, if discrete actions with 30 Hz frequency are used for supervision, the trajectory must first be decomposed into actions on a frame-basis and then used to train the model. Second, a smooth trajectory with high-frequency control certainly improves the success rate, which is demonstrated in the robotic community~\cite{kleff2021high}.

When early termination of \textsc{Corki}\xspace is applied, the accuracy trend initially increases and then decreases. This is because the shorter the length of the actual trajectory, the closer it aligns with discrete action supervision. However, if the trajectory taken by the robot is too long, useful environmental information may not be captured and utilized effectively, as the closed-loop feedback also operates at a lower frequency.

\underline{\textsc{Corki}\xspace-ADAP} works. This result validates our intuition that predicting a new trajectory whenever a significant movement occurs, such as a high curvature on the trajectory or a change in the status of the gripper, is beneficial.

%\yiyang{I'm not sure I'm right or not. I can think about three topics. The first one is why trajectory prediction much better? Trajectory prediction is generally more data-efficient because it can learn and adapt from continuous data streams. In contrast, discrete action prediction may require more action data for the neural network to learn these relationships. The second one is the trend of \textsc{Corki}\xspace-T,The difference between them lies in the granularity of the action space division. Considering one action per image is a common practice among AI algorithm designers, but it is not necessarily the optimal choice. If we divide the action space too finely, it may lead to excessive micro-operations, resulting in performance degradation. Therefore, 5 might be a better choice for the model's performance. The third one is the adaptive step algorithm achieves Pareto optimality, demonstrating the effectiveness of adaptive algorithms. However, it performs poorly on unseen tasks, which may be due to the correspondence of time steps in the test environment is not good. Therefore, new adaptive methods may be necessary to assist the network in learning. Maybe we can find more.}

\paragraph{Trajectory Comparison.} The accuracy of our applications is directly related to the correctness of the trajectory. Therefore, we provide detailed trajectory data for evaluation. We compare the error on the trajectory and show it in Fig.~\ref{fig:trajerror}. On average, \textsc{Corki}\xspace reduces the error by 25.0\%.

However, we have also observed that a lower trajectory error does not always correlate with higher accuracy. For instance, although \underline{\textsc{Corki}\xspace-3} has a lower mean trajectory error compared to \underline{\textsc{Corki}\xspace-5}, its success rate and average job length are lower. This discrepancy arises because the trajectory only reflects the trend of the robotic arm and cannot be treated as a perfect indicator of success rate. Additionally, this statistic does not account for the status of the gripper, which is also critical to the success of tasks.

% first column is the trajectory error,the second one is the maximun deviation of x,y,z
% P9t1 1.281608193788902 [0.2524349  0.19179267 0.28086687]
% P9t3 0.9948819075977617 [0.20355541 0.18481353 0.25035135]
% P9t5 1.1203152626443063 [0.23579604 0.21149901 0.25259639]
% P9t7 1.0996077538707896 [0.23440515 0.19730023 0.2382593 ]
% p9t9 1.128784922711947 [0.2316301  0.19419734 0.26122594]
% p9adaptive 1.1674543457989919 [0.25958707 0.20767254 0.2584243 ]
% baseline 1.5090353202556699 [0.29478786 0.242021   0.30303944]

We further illustrate the differences in trajectories with a real example. We compare trajectories on three dimensions separately and present the results in Fig.~\ref{fig:traj}. 

Although the baseline method can generate trajectories close to the ground truth on the Y dimension (Fig.~\ref{fig:trajy}) and Z dimension (Fig.~\ref{fig:trajz}), it clearly deviates from the target on the X dimension at time step 40 (Fig.~\ref{fig:trajx}). In contrast, \textsc{Corki}\xspace maintains alignment with the ground truth across all three dimensions. These results again emphasize that while trajectory is related to the success rate, it cannot fully determine task success. Even though \textsc{Corki}\xspace's trajectory slightly differs on the X dimension compared to the ground truth, it still successfully completes the task.

\subsection{Performance Comparison}
\label{sec:eval:perf}

\begin{figure}[t]
    \centering
    \includegraphics[width=1\columnwidth]{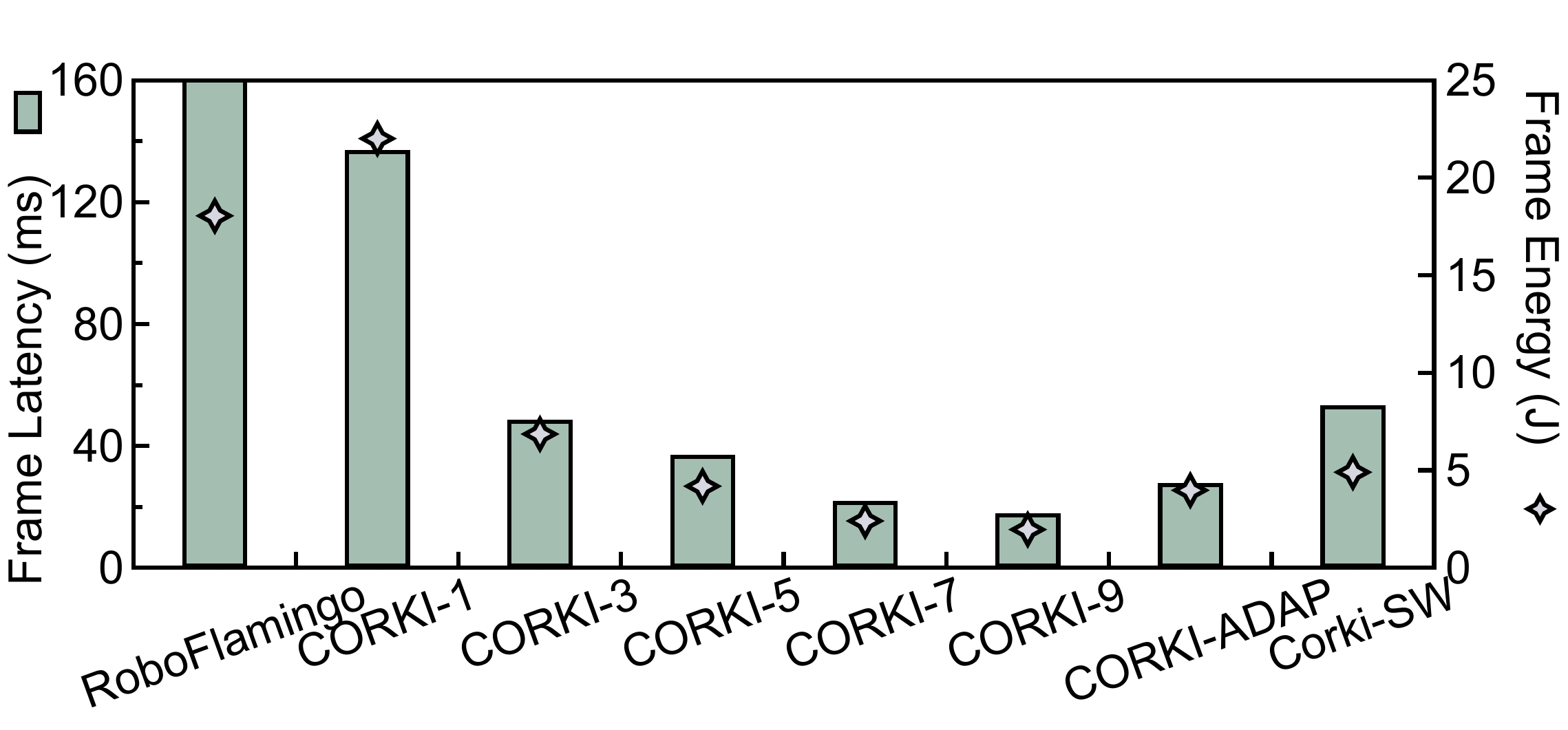}
    \caption{Runtime latency and energy consumption comparison between \textsc{Corki}\xspace and baselines. }
    \label{fig:performance}
    \vspace{-5pt}
\end{figure}

\paragraph{Latency Comparison.} We compare the latency and present the results in Fig.~\ref{fig:performance} on the left y-axis. \textsc{Corki}\xspace significantly reduces the frame latency of embodied AI robotic applications. Among the variations, \underline{\textsc{Corki}\xspace-9} achieves the best speedup of $9.1\times$, as the inference frequency of the large language model is reduced by $8\times$. As the length of the actual trajectory taken increases from 1 to 9, the speedup gradually increases from $1.2\times$ to $9.1\times$. On the other hand, \underline{\textsc{Corki}\xspace-ADAP} demonstrates a speedup of $5.9\times$, providing an ideal trade-off between accuracy and efficiency. Compared to \underline{\textsc{Corki}\xspace-5}, \underline{\textsc{Corki}\xspace-SW} has a 43.6\% longer average latency, as the actual control still happens on CPUs. Specifically, \underline{\textsc{Corki}\xspace-5} has an average 26.9 Hz frequency, while \underline{\textsc{Corki}\xspace-SW} only has 18.7 Hz. 

%We compare the latency and show the results in Fig.~\ref{fig:performance} on left y-axis. \textsc{Corki}\xspace significantly reduce the frame latency of embodied AI robotic applications. \texttt{\textsc{Corki}\xspace-9} has the best speed up of $3.6\times$ as the inference of large language model frequency is reduced by $8\times$. When the length of the actual trajectory taken increases from 1 to 9, the speed up gradually increases from $1.1\times$ to $3.6\times$. \texttt{\textsc{Corki}\xspace-ADAP} has a speed up of $3.0\times$, which provides an ideal trade-off between accuracy and efficiency. 

\paragraph{Energy Consumption Comparison.} \textsc{Corki}\xspace also significantly saves energy consumption. \underline{\textsc{Corki}\xspace-1} has slightly higher energy consumption compared to the baseline, as it takes one step for every predicted trajectory, which is similar to the baseline. Besides \underline{\textsc{Corki}\xspace-1}, all \textsc{Corki}\xspace variations have significantly lower energy consumption. \underline{\textsc{Corki}\xspace-9} has a $9.2\times$ energy reduction. Low energy consumption is critical to robots, which are mostly battery-supported devices.

\begin{figure*}[t]
%\vspace{-10pt}
\centering
\subfloat[\small{Per-frame latency breakdown}]
{
  \includegraphics[trim=0 0 0 0, clip, width=0.62\columnwidth]{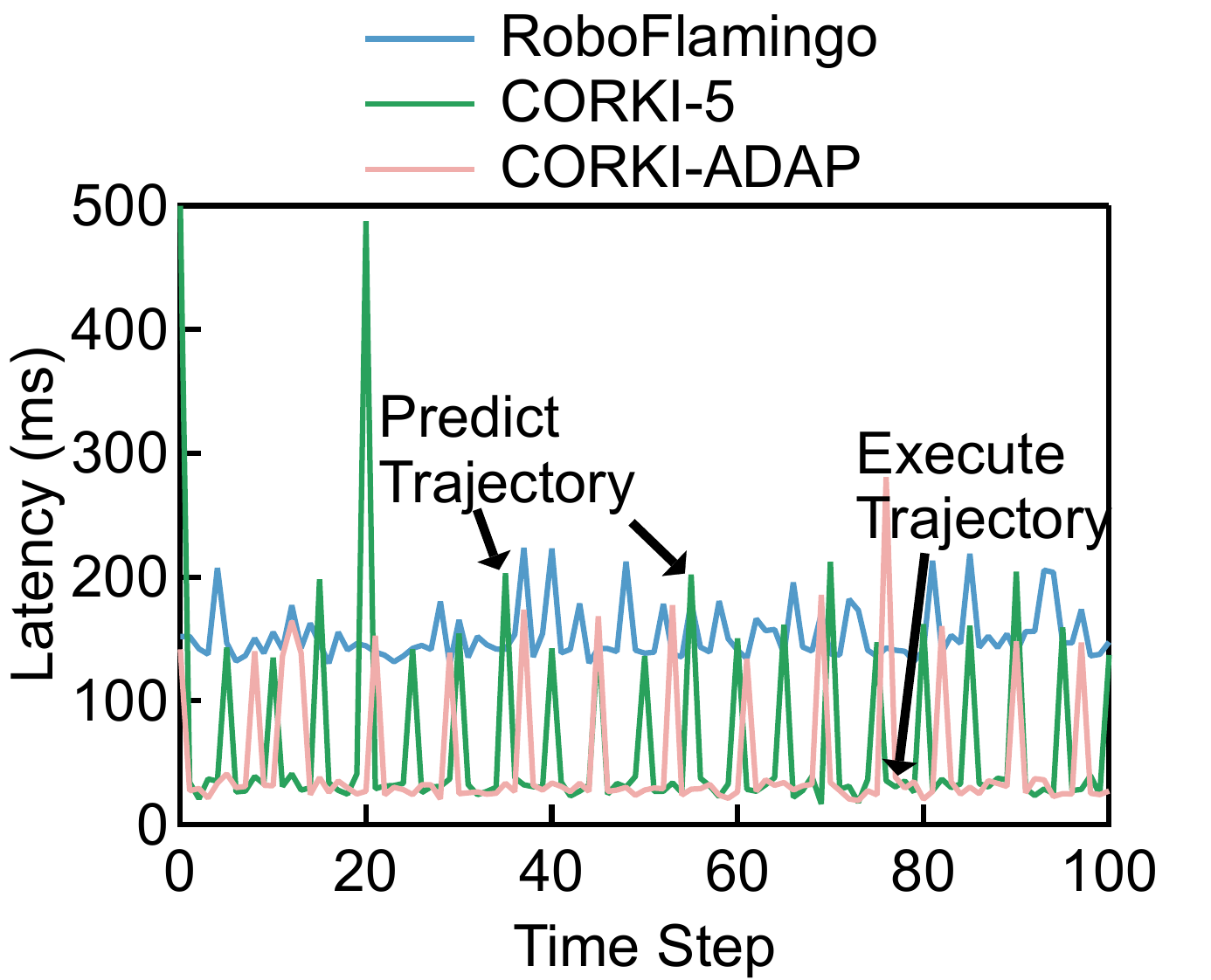}
  \label{fig:lat_break}
}
\subfloat[\small{Per-frame energy breakdown}]
{
  \includegraphics[trim=0 0 0 0, clip, width=0.62\columnwidth]{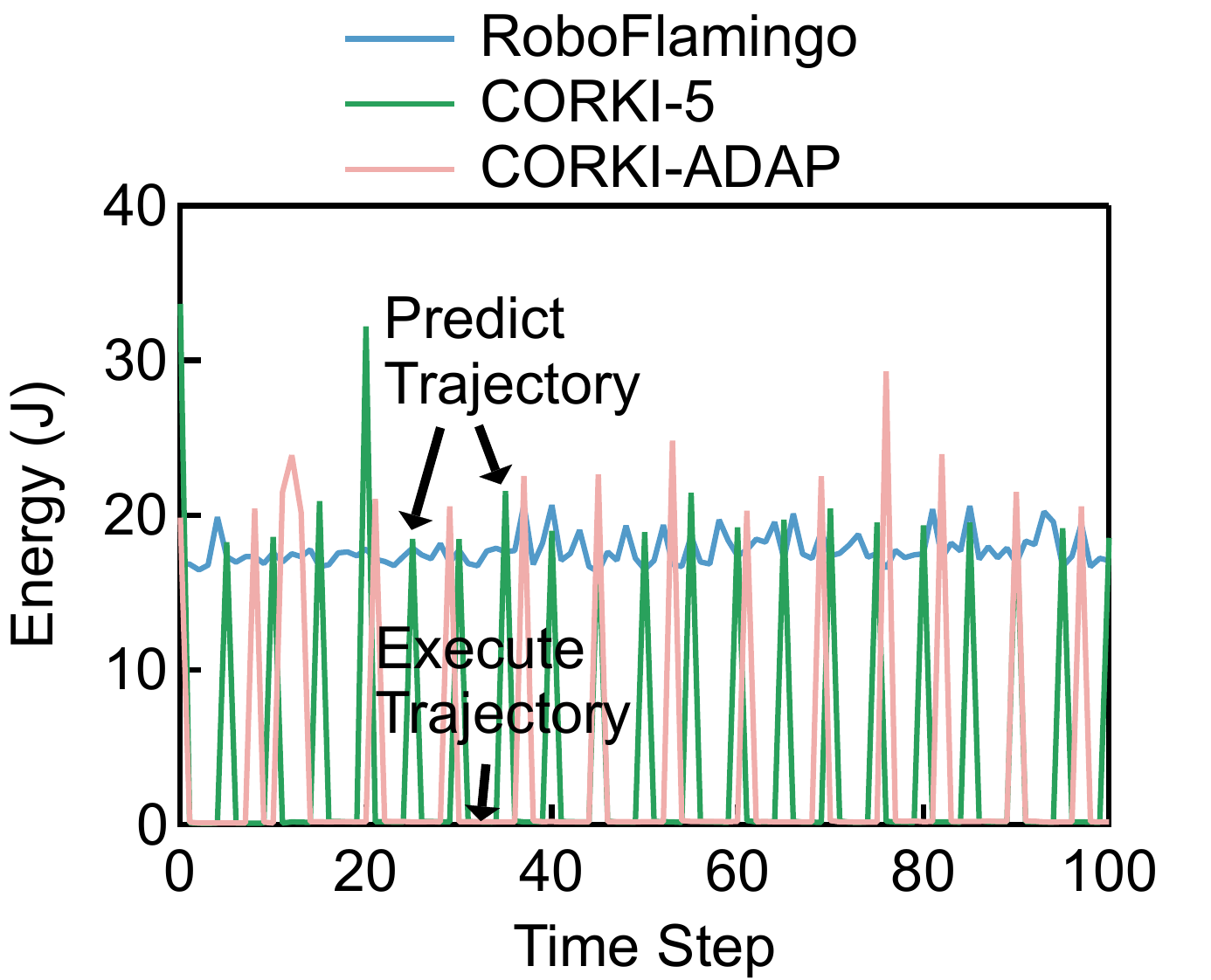}
  \label{fig:eng_break}
}
\subfloat[\small{Long tail problem and worst case latency analysis.}]
{
  \includegraphics[trim=0 0 0 0, clip, width=0.62\columnwidth]{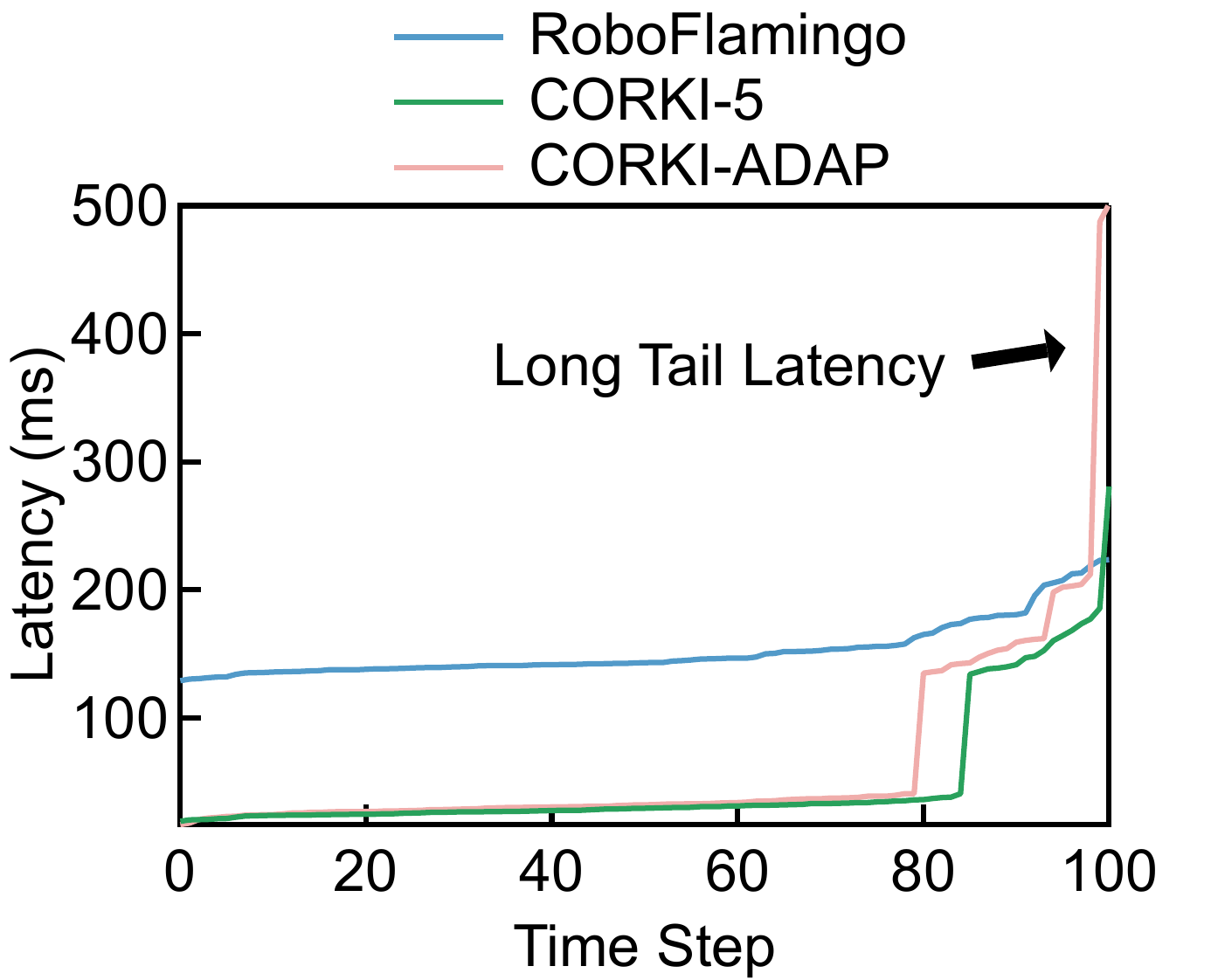}
  \label{fig:worst}
}
\vspace{-10pt}
\caption{Per-frame latency, energy comparison and long-tail analysis.}
\end{figure*}

\paragraph{Frame-by-frame Analysis.} We finally show a frame-by-frame analysis of latency and energy consumption for one single sequence. Fig.~\ref{fig:lat_break} shows the results of latency, and Fig.~\ref{fig:eng_break} shows the results for energy consumption. Both latency and energy consumption of \textsc{Corki}\xspace have the same trend, where the crest indicates the inference of LLM happening at that time step, and the trough means the robot is executing the trajectory predicted from the last time. \underline{\textsc{Corki}\xspace-5} has a periodical crest, as every 5-time steps, the inference will happen once. \underline{\textsc{Corki}\xspace-ADAP} has a more flexible crest and trough compared to \underline{\textsc{Corki}\xspace-5}. This is due to the waypoints identification and flexible length of the actual trajectory.

%\begin{figure}[t]
%    \centering
%    \includegraphics[width=0.8\columnwidth]{figs/worst.pdf}
%    \caption{\hl{Long tail problem and worst case latency analysis.}}
%    \label{fig:worst}
%\end{figure}

We find that although our method achieves lower average frame latency, it does exhibit severer long tail problem, as different frames undergo different execution pipelines. We sort the frame latency and show the results in Fig.~\ref{fig:worst}, result suggests the relative latency variation of the baseline is 56.0\% lower than \textsc{Corki}\xspace.

The acceleration comes from three sides. First, the inference frequency is largely reduced, which contributes to the most latency reduction. Second, \textsc{Corki}\xspace hardware successfully accelerates the control process by up to $29.0\times$, reducing the control latency. Finally, communication latency between the robot and the server is hidden as we enable pipelining. 

We vary the baselines in Tbl.~\ref{tbl:gpubase} and data representations in Tbl.~\ref{tbl:database} to better understand the performance. Our findings show that regardless of whether inference is performed on a state-of-the-art GPU like the H100 or an embedded GPU such as the Jetson Orin, \textsc{Corki}\xspace-ADAP consistently achieves high speedup (Jetson Orin shows longest inference latency, consuming over 0.9 second per frame, which makes it impossible to achieve real-time control). A similar conclusion holds when using different data representations.

\begin{table}[t]
\centering
\caption{Performance under different GPU/CPU baselines.}
\resizebox{1.0\columnwidth}{!}
{
\begin{tabular}{c|c|c|c|c}
\toprule[0.15em]
\textbf{GPUs} & \textbf{V100 (Currently used)} & \textbf{H100} & \textbf{Jetson Orin 32GB} & \textbf{Xeon 8260}\\
\midrule[0.05em]
\textbf{Normalized Inference Latency} & $1\times$ & $0.4\times$ & $10.0\times$ & $8.9\times$ \\
\textbf{Speedup} & $5.9\times$ & $6.4\times$ & $5.3\times$ & $5.4\times$  \\
\bottomrule[0.15em]
\end{tabular}
}
\label{tbl:gpubase}
\vspace{-10pt}
\end{table}

\begin{table}[t]
\centering
\caption{Performance under different data representations.}
\resizebox{1.0\columnwidth}{!}
{
\begin{tabular}{c|c|c|c}
\toprule[0.15em]
\textbf{GPUs} & \textbf{32-bit Float (Currently used)} & \textbf{16-bit Float} & \textbf{8-bit Int} \\
\midrule[0.05em]
\textbf{Normalized Inference Latency} & $1\times$ & $0.8\times$ & $0.4\times$ \\
\textbf{Speedup} & $5.9\times$ & $6.0\times$ & $6.4\times$ \\
\bottomrule[0.15em]
\end{tabular}
}
\label{tbl:database}
\vspace{-10pt}
\end{table}

\subsection{Sensitivity Study}

\begin{figure}[t]
%\vspace{-10pt}
\centering
\subfloat[\small{Normalized speedup increases when the level of approximation increases.}]
{
  \includegraphics[trim=0 0 0 0, clip, width=0.46\columnwidth]{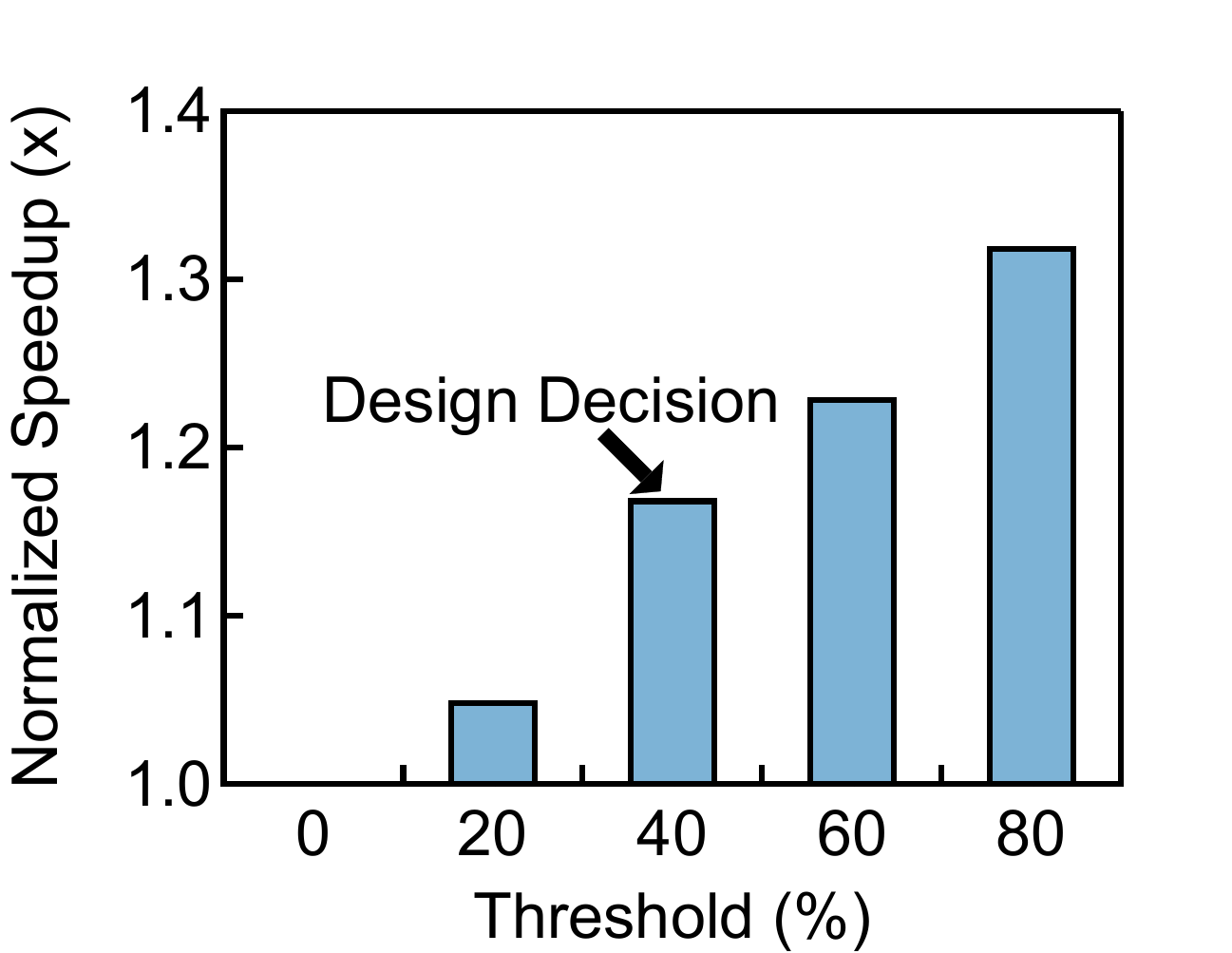}
  \label{fig:sensespeed}
}
\hfill
\subfloat[\small{Trajectory error increases when the level of approximation increases.}]
{
  \includegraphics[trim=0 0 0 0, clip, width=0.46\columnwidth]{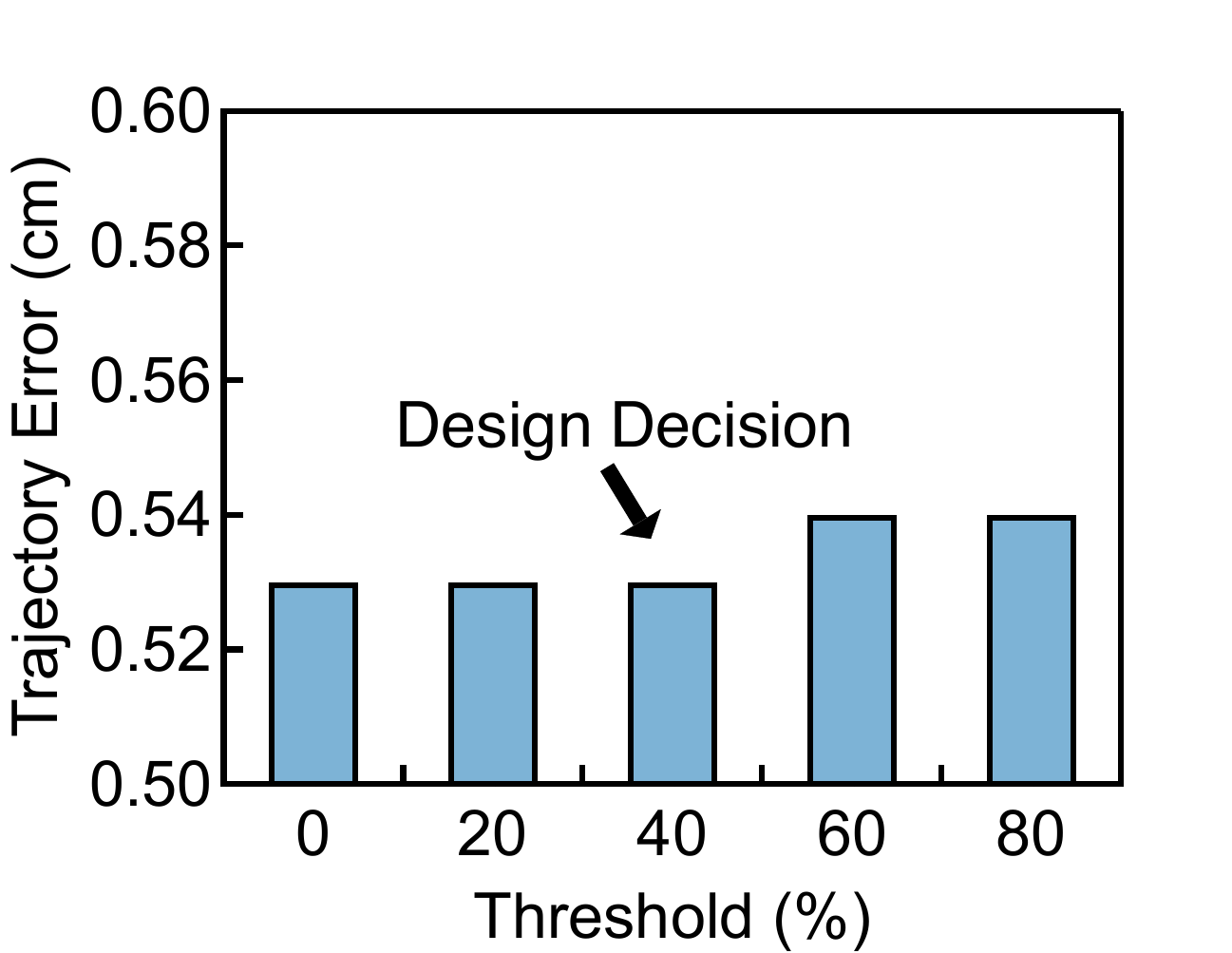}
  \label{fig:senseacc}
}
\caption{Relationship between speedup and trajectory error with respect to the approximation threshold.}
\label{fig:sense}
\vspace{-10pt}
\end{figure}

We further conduct experiments to analyze the impact of the approximation threshold on trajectory error and speedup within our control system. A higher threshold signifies a greater tendency to retain previously computed parameter values.  %The threshold represents the probability of applying approximate computing, where a higher threshold implies a greater likelihood of retaining previously computed parameters, thereby enabling more aggressive approximate computation.
Fig.~\ref{fig:sensespeed} illustrates the relationship between the speedup and the approximation threshold. As the threshold increases, approximate computation becomes more prevalent, leading to an increase in the speedup. Fig.~\ref{fig:senseacc} depicts the relationship between the trajectory error and the threshold. Across the range of thresholds, the trajectory error remains minimal. This can be attributed to the reduction in computation latency afforded by approximate computation, allowing for higher control frequencies and consequently, more robust control. In our design, we opt for the threshold of $40\%$ to balance speedup and accuracy.

\section{Related Work}
\label{sec:related}

\paragraph{Computing Systems for Embodied Artificial Intelligence.} Embodied Artificial Intelligence (EAI) differs from semantic AI by emphasizing agents, typically robots, that interact with the environment and execute long-horizon tasks. Recently, with the success of Large Language Models (LLMs) as planners, research in this domain has intensified, aiming to develop highly intelligent robots~\cite{duan2022survey,franklin1997autonomous,chrisley2003embodied,liu2024ok,vemprala2024chatgpt,huang2023voxposer}. While most studies focus on enhancing functionalities, our research emphasizes real-time performance. Our approach is rooted in the robotic community, where trajectory serves as the fundamental unit of planning and control. This contrasts with the predominant vision-centric perspective, which treats images or frames as the basic units.

\paragraph{Accelerators for Robotic Applications.} With the growing interest in treating robots as a new computing platform, our community has increasingly focused on dedicated accelerators for robotic computing. These accelerators have been designed for localization~\cite{suleiman2019navion,gan2021eudoxus,liu2021archytas,liu2019eslam,eyvazpour2023hardware,sugiura2022universal,liu2022mobilesp,sugiura2021unified,hao2022factor,liu2022energy}, motion planning~\cite{hsiao2023vapr,hao2023blitzcrank,huang2024moped,bakhshalipour2022racod,shah2023energy,lian2018dadu,murray2016microarchitecture,murray2019programmable}, control~\cite{neuman2021robomorphic, neuman2023roboshape,yang2023dadu,lian2017dadu,sacks2018robox,shao2018towards,aude1991hardware,gac2012fpga}, and more~\cite{mayoral2022robotcore,lienen2020reconros,hao2024orianna,lee2024spade,krishnan2022automatic,yu2020building,bakhshalipour2024tartan}. However, most accelerators focus on one or multiple modules within a traditional rule-based robotic computing system. Our work, in contrast, focuses on an end-to-end learning-based system, combining innovations in both algorithms and architecture, setting it apart from previous research.

\section{Discussion}
\label{sect:diss}
In this work, we focus on modifying the execution pipeline of embodied AI-powered robotic manipulation tasks, shifting from vision-centric frame-by-frame prediction to robot-centric trajectory prediction. Our results indicate that the proposed method performs well in a setting where a single robotic arm manipulates objects within a confined space, such as a desk.

Note that while the principle of predicting continuous trajectories instead of discrete actions can be extended to other robot types and tasks in the embodied AI domain, significant detailed design is required. At a minimum, we identify two key aspects to consider.

First, our method is limited to robotic arms, which typically have 9 DoF or fewer. Predicting the trajectory of a robot with a higher degree of freedom requires significantly greater effort. For instance, in the case of a humanoid robot, the trajectories of both the feet and arms must be predicted, while also considering their coordination. Simply applying our method to a humanoid robot may not work.

Second, our method can currently handle relatively long trajectories, given that sudden changes in the movement of a robotic arm are rare and that robotic arm's motion tends to be slow. However, in tasks where the robot moves quickly with abrupt changes, the trajectory prediction must adapt accordingly.

\paragraph{Safety concerns.} This work focuses on slowly moving, space-constrained collaborative robotic arms, which were initially introduced as a safer alternative as opposed to industrial robots~\cite{zanchettin2015safety}. Compared to our baseline, the safety concern of our approach is predicting a longer future trajectory. We try to mitigate this risk by incorporating closed-loop features. On the other hand, the higher control frequency and lower trajectory errors demonstrate that \textsc{Corki}\xspace enables a much smoother, jitter-less control compared to the baseline, thereby reducing safety concerns during execution.

\paragraph{End-to-end system power.} The power and energy saving reported in this paper pertain solely to the computing system. If we include the energy consumed by the motors powering the robots, the overall energy savings would be lower. In our setting, the computing system inside the robot accounts for 40.6\% of the total system power consumption (excluding server power).
\section{Conclusion}
\label{sec:concl}
Robots equipped with embodied AI algorithms often experience high latency due to the sequential execution pipeline and frequent LLM inference. In this paper, we propose \textsc{Corki}\xspace, a software-hardware co-design framework that significantly accelerates this process by transforming the algorithms to predict future trajectories, speeding up the control process, and pipelining communication with control. 
Results show that \textsc{Corki}\xspace achieves up to a $5.9\times$ speedup. \textsc{Corki}\xspace also achieves a maximum 13.9\% improvement in success rate.

% Results show that \textsc{Corki}\xspace achieves up to a $9.1\times$ speedup. \textsc{Corki}\xspace also achieves a maximum 17.3\% improvement in success rate.

% \vspace{-5pt}

\begin{acks}
We thank the anonymous reviewers for their valuable feedback.  This research was partially supported by the National Key Research and Development Program of China (Grant No. 2024YFB4505800), the Beijing Municipal Science and Technology Commission (Grant No. Z241100004224015) and the Longgang District Shenzhen's “Ten Action Plan” for Supporting Innovation Projects (under Grant LGKCSDPT2024002), whose support is gratefully acknowledged. Feng Yan and Lin Ma are the first author's mentors at Meituan.

\end{acks}

\bibliographystyle{ACM-Reference-Format}
\bibliography{refs}

\appendix

\end{document}